\documentclass[a4paper,12pt,twoside]{article}
\usepackage{amsmath}
\usepackage{graphicx}
\usepackage{graphics}

\def\rmi{\mathrm{i}} 
\def\rmd{\mathrm{d}} 
\def\rme{\mathrm{e}} 
\def\u{\:}           
\renewcommand{\vec}[1]{\mbox{\boldmath{$#1$}}} 
\newcommand{\bmi}[1]{\mbox{\boldmath{$#1$}}} 
\newcommand{\bvm}[1]{\mbox{\boldmath{$#1$}}} 
\newcommand{\boldsigma}{\mbox{\boldmath$\sigma$}} 
\newcommand{\boldphi}{\mbox{\boldmath$\phi$}} 
\usepackage{amsfonts}
 \DeclareSymbolFont{upgreek}{U}{eur}{m}{n}
 \DeclareMathSymbol{\rmalpha}{0}{upgreek}{"0B}
 \DeclareMathSymbol{\rmbeta}{0}{upgreek}{"0C}
 \DeclareMathSymbol{\rmgamma}{0}{upgreek}{"0D}
 \DeclareMathSymbol{\rmdelta}{0}{upgreek}{"0E}
 \DeclareMathSymbol{\rmmu}{0}{upgreek}{"16}
 \DeclareMathSymbol{\rmpi}{0}{upgreek}{"19}
 \DeclareMathSymbol{\rmsigma}{0}{upgreek}{"1B}
 \DeclareMathSymbol{\rmphi}{0}{upgreek}{"1E}
 \DeclareMathSymbol{\rmomega}{0}{upgreek}{"21}

\newcommand{\av}[1]{\left\langle#1\right\rangle}

\begin{document}

\title{Beam Extraction and Transport}

\author{T.~Kalvas\\
Department of Physics, University of Jyv\"askyl\"a,\\
40500 Jyv\"askyl\"a, Finland}

\maketitle

\begin{abstract}
This chapter gives an introduction to low-energy beam transport systems,
and discusses the typically used magnetostatic elements (solenoid, dipoles
and quadrupoles) and electrostatic elements (einzel lens, dipoles and
quadrupoles). The ion beam emittance, beam space-charge effects and the
physics of ion source extraction are introduced. Typical computer
codes for analysing and designing ion optical systems are mentioned,
and the trajectory tracking method most often used for extraction
simulations is described in more detail.
\end{abstract}

\section{Introduction}

\begin{figure}[htbp!]
  \centering
  \includegraphics[width=9cm]{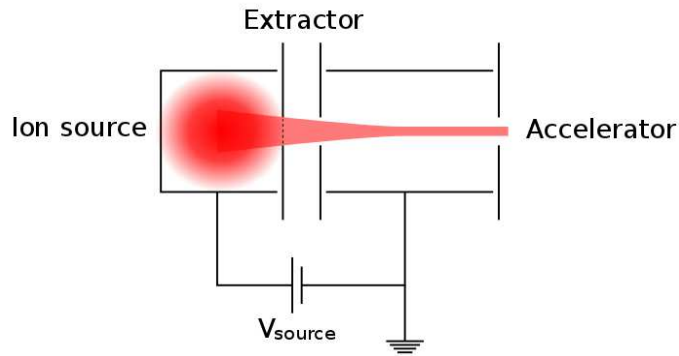}
  \caption{The most basic electrostatic extraction system possible}
  \label{fig:concept}
\end{figure}

In principle, the task of beam extraction and the following low-energy
beam transport (LEBT) system are quite simple. The ion source
extraction consists of the front plate of the ion source, which is
known as the plasma electrode, and at least one other electrode, the
puller (or extractor) electrode, which provides the electric field for
accelerating the charged particles from the ion source to form an ion
beam. Whether or not the extraction contains any other electrodes, the
beam leaves the extraction at energy
\begin{equation}
  E = q(V_{\text{source}}-V_{\text{beamline}}),
\end{equation}
defined by the charge $q$ of the particles and the potential
difference between the ion source, $V_{\text{source}}$, and the
following beamline, $V_{\text{beamline}}$, which is typically the
laboratory ground, as shown in Fig.~\ref{fig:concept}. The ion
source voltage is therefore set according to the requirements of the
subsequent application. The intensity of the particle beam depends, as
a first approximation, on the flux of charged particles hitting the
plasma electrode aperture. The extracted ion beam current can
therefore be estimated as
\begin{equation}
  I = \tfrac{1}{4} Aqn\bar{v},
\end{equation}
where $A$ is the plasma electrode aperture, $q$ is the charge of the
particles, $n$ is the ion density in the plasma and $\bar{v}$ is the
mean velocity of extracted particles in the ion source
plasma. Assuming a Maxwell--Boltzmann distribution for the extracted
plasma particles, the mean velocity $\bar{v}=\sqrt{8kT/\rmpi m}$. From
the point of view of the extraction, the plasma electrode aperture can be
adjusted to tune the beam intensity.

The practical solutions are unfortunately much more complicated in
most cases. The applications following the LEBT,
which typically are accelerators to bring the beam to higher energies,
often pose strict requirements for the ion beam parameters. Not only
do the beam intensity, energy and species spectrum need to meet the
requirements, but also the beam spatial and temporal structure are
specified. For an example of what kind of parameters the
specifications might contain, please see Table~\ref{tab:specifications}.

\begin{table}[htbp!]
  \begin{center}
    \caption{Specifications of the H$^-$ ion beam at the end of the
      LEBT, entering the 2.5{\u}MeV Radio Frequency Quadrupole (RFQ)
      accelerator of the Spallation Neutron Source (SNS) \cite{Thomae2002}.}
    \label{tab:specifications}
    \begin{tabular}{p{6cm}cc}
      \hline\hline
      \textbf{Parameter}            & \textbf{Value} \\
      \hline
      \noalign{\vspace{2pt}}
      Beam current (H$^-$)          & 50{\u}mA \\
      Beam energy                   & 65{\u}keV \\
      Emittance (normalized r.m.s.) & 0.20{\u}mm{\u}mrad \\
      Twiss $\alpha$                & 1.5 \\
      Twiss $\beta$                 & 0.06{\u}mm{\u}mrad$^{-1}$ \\
      Macro-pulse length            & 1{\u}ms \\
      Macro-pulse duty factor       & 6\% \\
      Mini-pulse length             & 645{\u}ns \\
      Mini-pulse duty factor        & 68\% \\
      \hline\hline
    \end{tabular}
\end{center}
\end{table}

The spatial requirements for the beam mean that focusing is necessary
in the LEBT. Similarly, the temporal requirements necessitate beam
chopping.  Without careful design of the focusing elements, the space-charge force of the beam blows up the beam to the walls of the vacuum
chamber, and only a part of the generated beam gets transported to the
following accelerator. The extraction focusing systems must also
provide some adjustability because, in most cases, the plasma conditions
might not be constant in day-to-day operations. The LEBT has to be able
to adapt to days of lower and higher performance, while maximizing the throughput
to the following accelerator. Designing such systems is not easy. From
the ion optics point of view, a system that is as short as possible would be
preferred, but at the same time the system design also has to take into
account the practical engineering constraints. For example, the beamline needs to have space for diagnostics and vacuum pumps in addition
to the focusing elements.

This chapter concentrates mainly on the topic of ion optics in
low-beam-energy systems: the focusing elements, beam space-charge blow-up,
beam--plasma interface and how to model these systems with computer codes.

\section{Low-energy beam transport}

The ion beam travels in the beam transport line from one ion optical
element to another along a curved path, which is usually defined as
the longitudinal direction $z$. The transverse directions $x$ and $y$
are defined relative to the centre of the transport line, the optical
axis, where $x=0$ and $y=0$. The transport line is usually designed in
such a way that a so-called reference particle travels along the
optical axis with nominal design parameters. The ion beam (bunch) is
an ensemble of charged particles around the reference particle, with
each individual particle at any given time described by spatial
coordinates $(x,y,z)$ and momentum coordinates $(p_x,p_y,p_z)$. This
six-dimensional space is known as the particle \textit{phase space}. In
addition to these coordinates, often inclination angles $\alpha$ and
$\beta$ or the corresponding tangents $x'$ and $y'$ are used. These
are defined by
\begin{equation}
  x'= \tan \alpha = \frac{p_x}{p_z}
    \quad \text{and} \quad
  y' = \tan \beta = \frac{p_y}{p_z}.
\end{equation}

The motion of a charged particle in electromagnetic fields $E$ and $B$
is described by the Lorentz force $F$ and Newton's second law, giving
\begin{equation}
  \frac{\rmd\vec{p}}{\rmd t} = \vec{F} = q(\vec{E} + \vec{v} \times \vec{B}),
  \label{eq:motion}
\end{equation}
where $p$ is the momentum and $v$ is the velocity of the particle with
charge $q$. In general, the trajectory of a charged particle can be
calculated by integrating the equation of motion if the fields are
known. In the case of beam transport, the fields have two origins:
(i)~external fields, which are mainly generated by the ion
optical elements; and (2)~beam-generated fields. Generally, in
LEBT ion optics, we assume that the particle density in a beam is low
enough that single particle--particle interactions are negligible. It
is therefore sufficient to take into account the collective beam-generated fields.

\subsection{Beamline elements}

The ion optical elements of the beam transport line come in two
varieties: magnetic and electric. In the case of high-energy beams,
where $v \approx c$, magnetic elements are used because the force,
which is created with an easily produced magnetic field of 1{\u}T,
equals the force from an electric field of 300{\u}MV{\u}m$^{-1}$, which is
impossible to produce in a practical device. In LEBT systems, where
the beam velocity is low, and the practical limit for electric fields
is about 5{\u}MV{\u}m$^{-1}$, the achievable forces are comparable, and other
factors, such as size, cost, power consumption and the effects of beam
space-charge compensation, come into play. An important factor in the
selection of the type of beamline elements is also the fact that
electrostatic fields do not separate ion species. In electrostatic
systems the particles follow trajectories defined only by the system
voltages. The magnetic elements, on the other hand, have a dependence on
mass-to-charge ratio $m/q$. This allows separation of different
particle species from each other.

The common beamline elements that are used to build LEBT
systems include immersion lens, einzel lens, solenoid, dipole and
quadrupole lenses. A short introduction to each of these elements is
given below. For more detailed analyses, the reader is referred to
literature (Refs.~\cite{Liebl2008} and \cite{Wollnik1987}, for example).

\subsubsection{Immersion lens}

The immersion lens (or gap lens) is simply a system of two electrodes
with a potential difference of $\Delta V = V_2 - V_1$. The lens can be
either accelerating or decelerating, and, in addition to changing the
particle energy by $q \Delta V$, the element also has a focusing
action. The focal length of the immersion lens is given by \cite{Liebl2008}
\begin{equation}
  \frac{f}{L} = \frac{4(\sqrt{V_2/V_1}+1)}{V_1/V_2+V_2/V_1-2},
\end{equation}
where $L$ is the distance between the electrodes. The electrostatic
extraction systems always have gap lenses, which accelerate the beam
to the required energy. The first acceleration gap (plasma electrode
to puller electrode) is a special case of the immersion lens because
of the effect of the plasma on the electric field. It will be
addressed later in this chapter.

\subsubsection{Einzel lens}

\begin{figure}[htbp!]
  \centering
  \includegraphics[width=5.0cm]{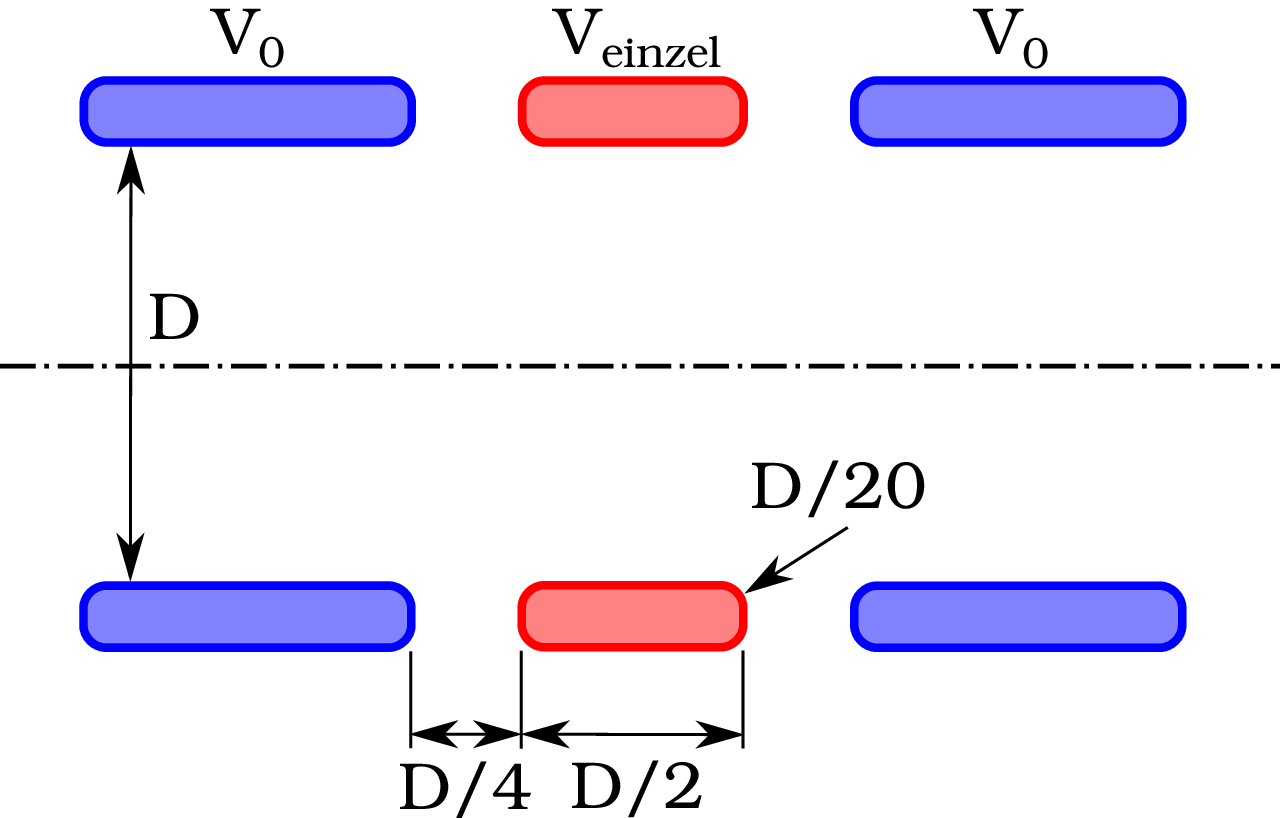}
  \includegraphics[width=9.0cm]{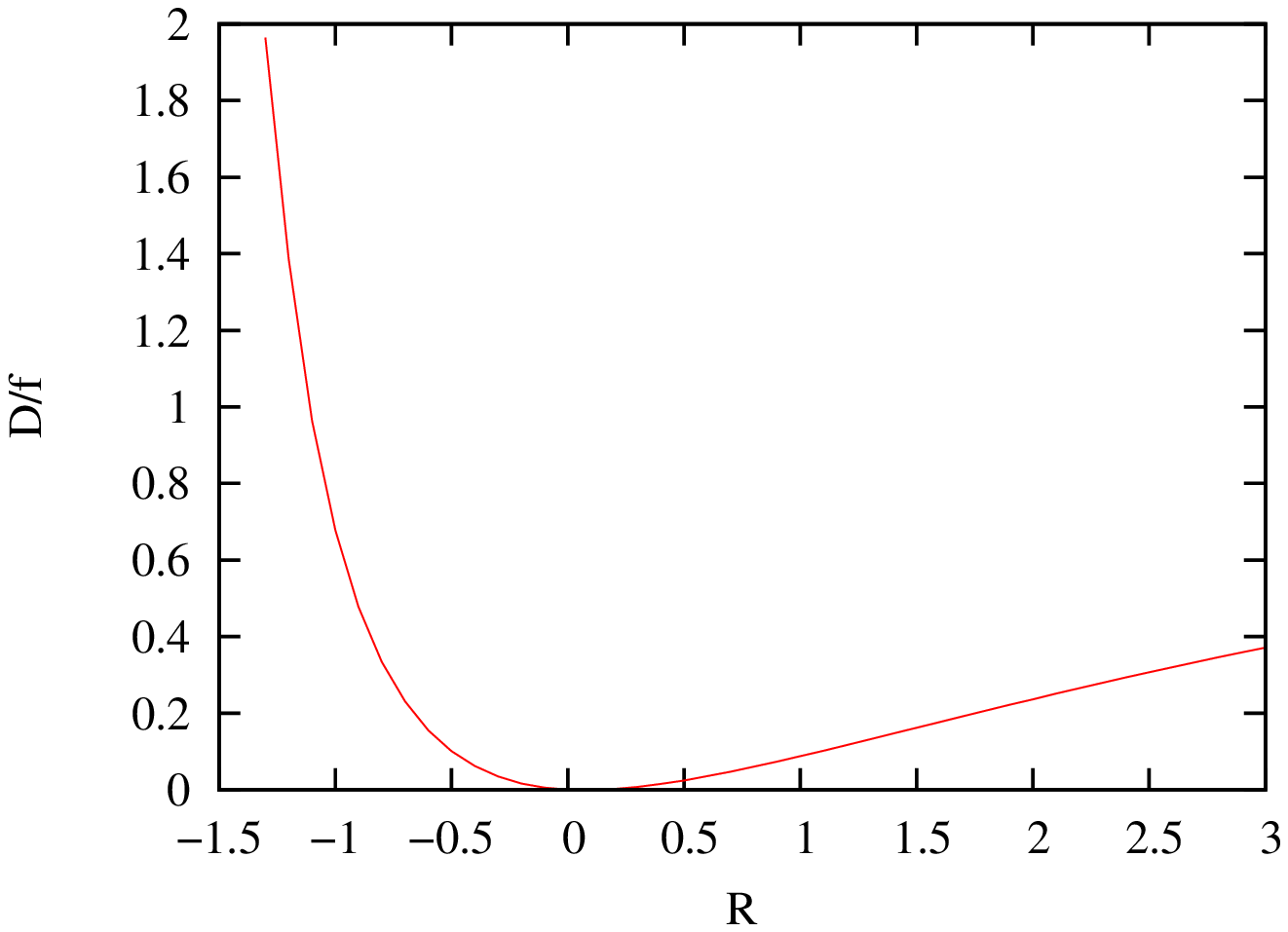}
  \caption{An example geometry for an einzel lens and its refractive
    power scaled with the einzel internal diameter $D$ as a function
    of the voltage ratio $R$. The lens is much stronger in
    decelerating mode compared to accelerating mode.}
  \label{fig:einzel_power}
\end{figure}

The einzel lens is made by combining two gap lenses into one
three-electrode system with first and last electrodes at the beamline
potential $V_0$ and the centre electrode at a different potential
$V_\text{einzel}$. The einzel lens, which is typically cylindrically
symmetric for round beams, is the main tool for beam focusing in many
electrostatic extraction systems. The einzel focusing power is
dependent on the geometry and the voltage ratio $R =
(V_\text{einzel}-V_0)/V_0$, assuming that zero potential is where
the beam kinetic energy is zero. The einzel lens may have the first
gap accelerating and the second gap decelerating (known as accelerating
einzel lens, $R>0$) or vice versa (known as decelerating einzel lens,
$R<0$). Both configurations are focusing, but the refractive power of
the einzel in decelerating mode is much higher than in accelerating
mode with the same lens voltage. For example, for the geometry shown in
Fig.~\ref{fig:einzel_power}, the focal length $f\approx 10 D$ for a
decelerating mode, $V_\text{einzel}-V_0=-0.5V_0$. To achieve the same focal
length in accelerating mode, a voltage $V_\text{einzel}-V_0=1.1V_0$ is
needed. On the other hand, accelerating einzel lenses should be
preferred if the required higher voltage (and electric fields) can be
handled, because they have lower spherical aberrations than
decelerating einzel lenses, especially when the required refractive
power is high.

A special case of the einzel lens, where the first electrode and the
third electrode are at different potentials, is also possible. This
kind of set-up is known as a three-aperture immersion lens or
zoom lens. It provides adjustable focusing in a system that otherwise
acts as an immersions lens \cite{Liebl2008}.

\subsubsection{Solenoid lens}

A solenoid lens is the magnetic equivalent of the electrostatic einzel
lens. It consists of rotationally symmetric coils wound around the
beam tube, creating a longitudinal magnetic field peaking at the
centre of the solenoid. The focusing action of the solenoid is
somewhat difficult to derive, but the idea can be described as follows,
assuming a thin lens \cite{Kumar09}. The radial magnetic field at the
entrance of the solenoid gives the particle entering the field with
$v_r = 0$ at radius $r_0$ an azimuthal thrust
\begin{equation}
  v_\theta = \frac{qBr_0}{2m},
  \label{eq_sol_azimuthal_thurst}
\end{equation}
which makes the trajectories helical inside the solenoid. At the exit
of the solenoid, the particle receives a thrust cancelling the
azimuthal velocity, but leaving the particle with a radial velocity
\begin{equation}
  v_r = -\frac{r_0q^2}{4m^2v_z} \int B^2 \, \rmd z.
  \label{eq_sol_radial_thurst}
\end{equation}
This radial velocity causes the beam to converge towards the optical
axis. The refractive power of the lens is given by
\begin{equation}
  \frac{1}{f} = \frac{q^2}{8mE} \int B^2 \, \rmd z.
\end{equation}

\subsubsection{Electrostatic and magnetic dipoles}

The electrostatic dipole and magnetic dipole are elements that are
primarily used to deflect charged-particle beams. The magnetic dipole is
constructed from coil windings, creating a constant magnetic field in
the transverse direction. The particles in the magnetic field follow
circular trajectories as usual, with radius
\begin{equation}
  \rho = \frac{p}{qB} \approx \frac{mv_z}{qB} =
  \frac{1}{B} \sqrt{\frac{2mV_\text{0}}{q}},
\end{equation}
where $V_\text{0}$ is the voltage used to accelerate the particles
from zero to $v_z$. Similarly, an electrostatic dipole may be
constructed from cylindrical electrodes of radii $r_1$ and $r_2$ with
voltages $V_1$ and $V_2$. The radius of curvature of the particle
between the plates becomes
\begin{equation}
  \rho = \frac{2 V_0}{E},
\end{equation}
where $E$ is the electric field and $V_0$ is the potential at the
orbit (again assuming that zero potential is where the beam kinetic
energy is zero). The voltage and electric field between the plates are
\begin{eqnarray}
  V &=& V_1 + (V_2-V_1) \frac{\log(r/r_1)}{\log(r_2/r_1)}, \\
  E &=& - \frac{V_2 - V_1}{\log(r_2/r_1)} \, \frac{1}{r}.
\end{eqnarray}
By choosing the plate voltages symmetrically as
$V_1=V_0+V_\text{plate}$ and $V_2=V_0-V_\text{plate}$, the required
plate voltage can be found as
\begin{equation}
  V_\text{plate} = V_0 \log(r_2/r_1).
\end{equation}
The optical axis of such a system is at radius $\rho=\sqrt{r_1
  r_2}$. This is not the only possibility for the cylindrical
dipole. The optical axis can also be chosen to be in the middle of the
plates, which leads to asymmetric voltages.

\begin{figure}[htbp!]
  \centering
  \includegraphics[width=6.5cm]{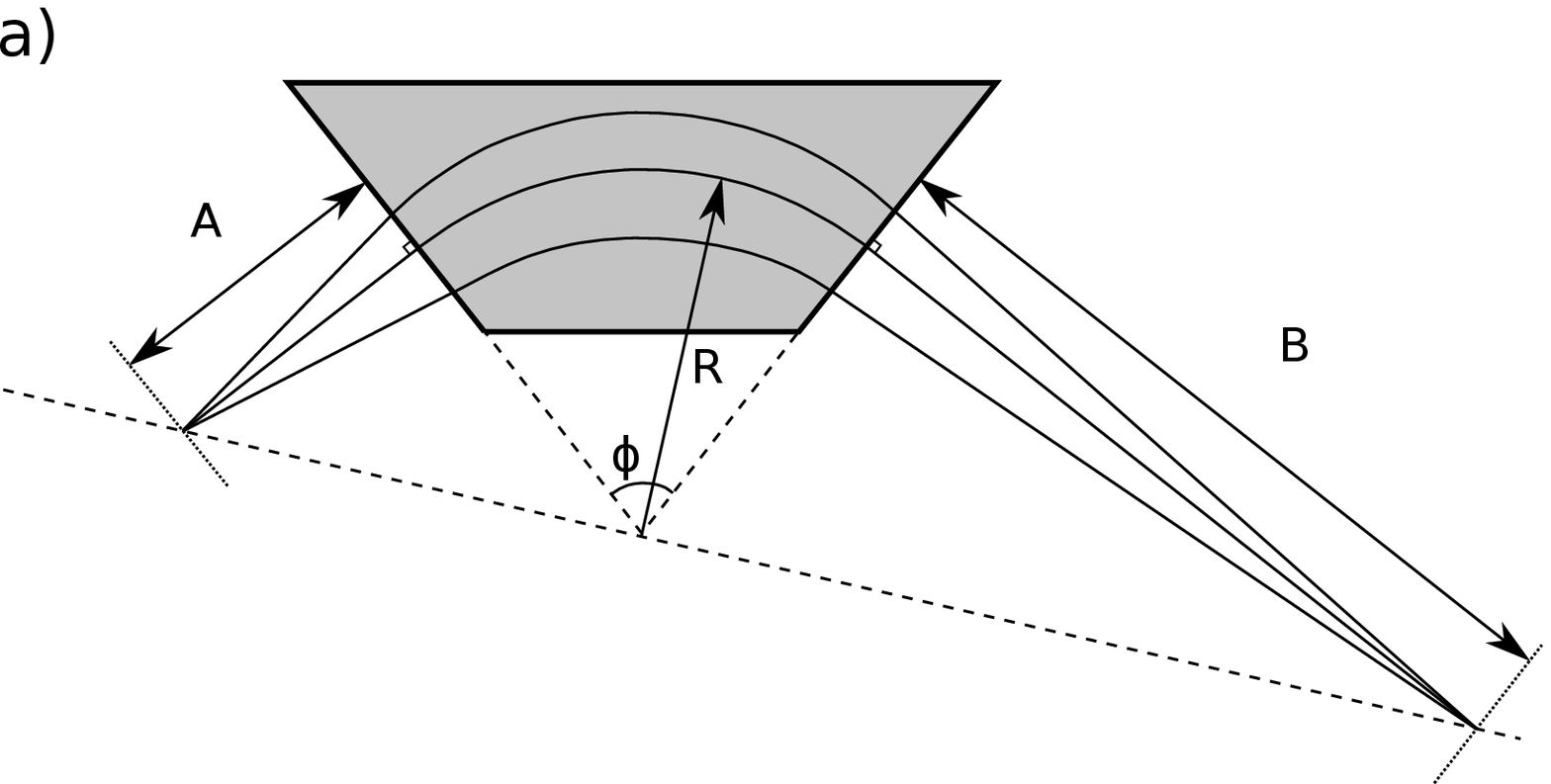}
  \hspace*{8mm}
  \includegraphics[width=8.0cm]{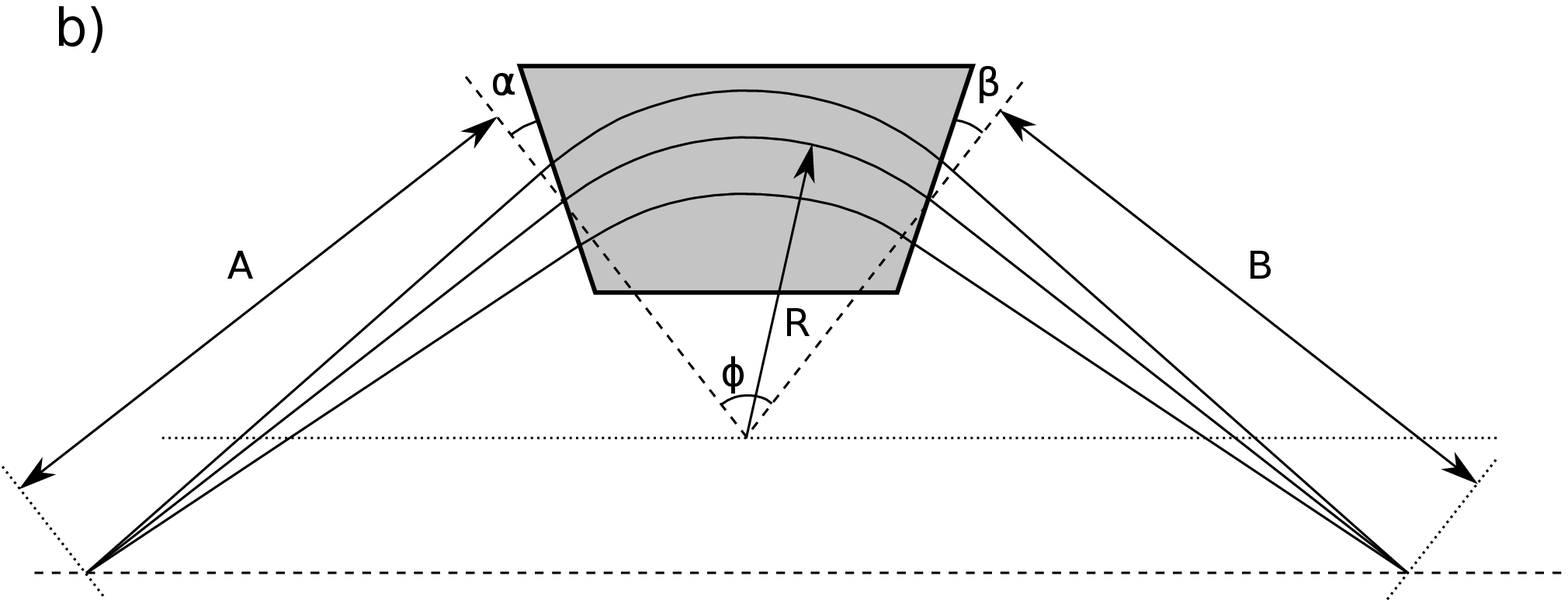}
  \caption{Focusing of a magnetic dipole
    in the bending plane. (a)~The case where the magnet has 0$^\circ$ edge
    angles can be described by Barber's rule: the center of curvature of
    optical axis and the two focal points are on a straight line. (b)~If
    the edge angles are positive, as shown, the focusing power is
    decreased.}
  \label{fig:mag_dip_focusing1}
\end{figure}

The dipole elements also have focusing/defocusing properties. 
For example, the magnetic dipole with edges 
perpendicular to the optical axis (edge angle 0$^\circ$) focuses the 
beam in the bending plane ($x$) as shown in Fig.~\ref{fig:mag_dip_focusing1}(a).
Directly from geometry, a so-called
Barber's rule can be derived: the centre of curvature of the optical
axis and the two focal points are on a straight line. For a symmetric
set-up this means that $A=B=R/\tan({\phi}/{2})$. There is no focusing
action in the $y$ direction.

\begin{figure}[htbp!]
  \centering
  \includegraphics[width=9.0cm]{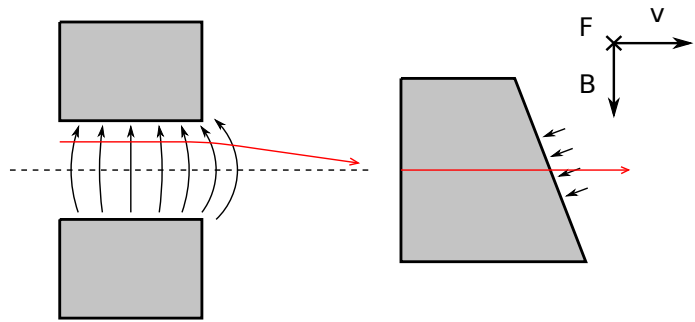}
  \caption{In a dipole magnet with positive edge angle, the fringing
    field has a $B_x$ component at non-zero $y$ coordinates, focusing
    the beam in the $y$ direction.}
  \label{fig:mag_dip_focusing2}
\end{figure}

If the magnet edge angles deviate from $90^\circ$, the
focusing power in the $x$ direction can be adjusted. If the edge angle is
made positive (as shown in Fig.~\ref{fig:mag_dip_focusing1}(b)), there
is weaker focusing in the $x$ direction. If the angle is negative, there
is stronger focusing in the $x$ direction. Changing the edge angle
also has an important effect in the $y$ direction: if the angles are
positive, the fringing field of the magnet will focus the beam in the
$y$ direction, as shown in Fig.~\ref{fig:mag_dip_focusing2}. Overall,
this means that the focusing in the $x$ direction can be traded for
$y$ focusing. The focal length from the edge focusing is given by
\begin{equation}
 f_y=\frac{R}{\tan \alpha}.
\end{equation}
In a symmetric double-focusing dipole (where there is the same focal
length in $x$ and $y$ directions), the angles and distances are given by
\begin{eqnarray}
  & 2 \tan \alpha = 2 \tan \beta = \tan({\phi}/{2}) , \\
  & A = B = \displaystyle\frac{2R}{\tan({\phi}/{2})} .
\end{eqnarray}
For a $\phi=90^{\circ}$ bending magnet, the edge angles become
$\alpha=\beta=26.6^{\circ}$ and the focal distances $A=B=2R$.

\begin{figure}[htbp!]
  \centering
  \includegraphics[width=7.5cm]{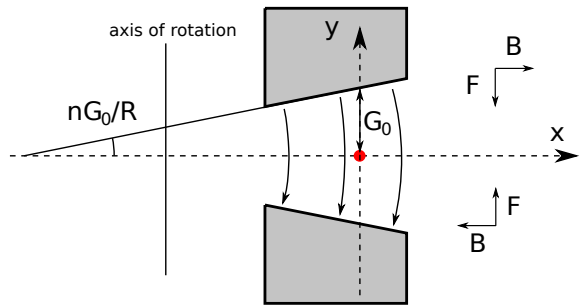}
  \caption{In the inhomogeneous sector magnets, the field strength
    decreases with increasing radius $x$ due to increasing gap, which
    also leads to $B_x$ increasing with $y$.}
  \label{fig:mag_dip_inhomo}
\end{figure}

A dipole magnet focusing in the $y$ direction can also be made by
making the pole faces conical, as shown in Fig.~\ref{fig:mag_dip_inhomo}.
This kind of magnet is known as a radially
inhomogeneous sector magnet. The magnetic field of such a magnet can be
approximated as
\begin{eqnarray*}
  B_y(x,y) &=& B_0 \left( 1 - n\frac{x}{R} + \cdots \right),\\
  B_x(x,y) &=& B_0 \left( n\frac{y}{R} + \cdots \right),
\end{eqnarray*}
where $B_0$ is the magnetic field on the optical axis with radius $R$
and $n$ is the so-called field index, which depends on the angle of the
magnet pole. The first-order approximation shows that the magnet is
focusing in the $x$ direction if $n<1$ and focusing in the $y$ direction if
$n>0$. The focusing forces inside the magnet are symmetric at
$n=\tfrac{1}{2}$.

The regular cylindrical electrostatic dipole only has focusing in the
$x$ direction similar to the regular straight edge magnetic dipole. The
$y$ focusing can be introduced by adjusting the ends of the cylindrical
plates for edge focusing or by using spherical or toroidal plates.

\begin{figure}[htbp!]
  \centering
  \includegraphics[width=0.70\textwidth]{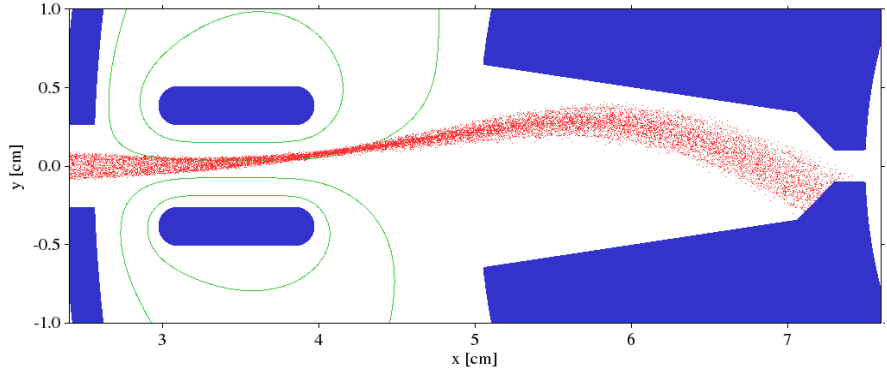}
  \caption{Fast beam chopping can be done with parallel plates. A
    particle-in-cell (PIC) simulation of an LBNL-built neutron
    generator using 15{\u}ns risetime $\pm 1500${\u}V switches for
    generating 5{\u}ns beam pulses is shown.}
  \label{fig:chop}
\end{figure}

For small-angle deflection, typically electrostatic parallel plates or
so-called $XY$ magnets are used. The parallel plates with $\pm
V_\text{plate}$ voltages bend the beam by an angle
\begin{equation}
  \theta = \frac{V_\text{plate}L}{V_\text{0}d},
\end{equation}
where $L$ is the length of the plates in the $z$ direction and $d$ is the
distance between the plates. This kind of system is typically used for
small corrections in beamlines and for beam chopping. For example, in
Fig.~\ref{fig:chop}, a simulation of a fast chopping system in an
LBNL-built neutron generator is shown. The $XY$ magnets are the magnetic
equivalent of parallel plates, with typically two pairs of windings
in a single instrument for correction in both transverse
directions. The beam deflection is given by
\begin{equation}
  \theta = LB\sqrt{\frac{q}{2mV_\text{0}}},
\end{equation}
where $L$ is the field length and $B$ is the field strength inside the device.

\subsubsection{Quadrupole lenses}

\begin{figure}[htbp!]
  \centering
  \includegraphics[width=6.5cm]{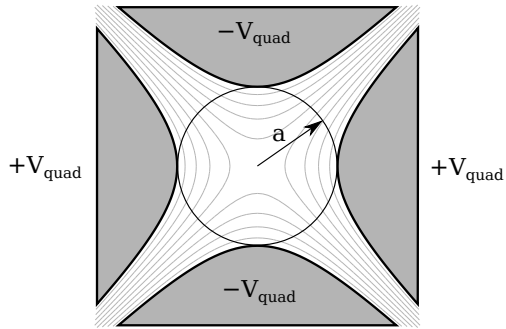}
  \caption{Electrostatic quadrupole, with $a$ being the distance from
    the optical axis to the electrode tip}
  \label{fig:quad}
\end{figure}

Electrostatic and magnetic quadrupoles are often used as focusing
elements in LEBT systems in addition to einzel lenses and solenoids.
The electrostatic quadrupole consists of four hyperbolic electrodes
placed symmetrically around the beam axis with positive potential
$V_\text{quad}$ on the electrodes in the $+x$ and $-x$ directions and
negative potential $-V_\text{quad}$ on the electrodes in the $+y$ and
$-y$ directions, as shown in Fig.~\ref{fig:quad}. The potential in
such a configuration is given by
\begin{equation}
  V = \frac{x^2-y^2}{a^2} V_\text{quad},
\end{equation}
where $a$ is the radius of the quadrupole. This leads to an electrostatic
field
\begin{equation}
  \vec{E} = -\frac{2V_\text{quad}}{a^2}x \bmi{\hat{x}} +
  \frac{2V_\text{quad}}{a^2}y \bmi{\hat{y}},
\end{equation}
from which we can see that such a quadrupole focuses a positive ion beam
in the $x$ direction and defocuses in the $y$ direction. By analysing the
particle trajectories in such fields, it can be shown that the
refractive powers of the system are
\begin{eqnarray}
  1/f_x &=& k \sin(kL),
  \label{eq:quad_focus_x}
  \\
  1/f_y &=& -k \sinh(kL),
  \label{eq:quad_focus_y}
\end{eqnarray}
where $k^2={V_\text{quad}}/{a V_0}$ and $L$ is the effective length
of the quadrupole.

The magnetic quadrupole has a similar construction to the
electrostatic one: the magnet poles are made to be hyperbolic and
windings are coiled in such a way that every other pole has magnetic
flux in to the beam and every other out of the beam. The magnetic
field in such a system is
\begin{equation}
  \vec{B} = \frac{B_{\rm T}}{a}y \bmi{\hat{x}} + \frac{B_{\rm T}}{a}x \bmi{\hat{y}},
\end{equation}
where $B_{\rm T}$ is the magnetic field density at the pole tip. Positively
charged particles having velocity $\vec{v}=v_z\bmi{\hat{z}}$ feel a force
$\vec{F}=qB_Tv_z(-x\bmi{\hat{x}}+y\bmi{\hat{y}})/a$, which is focusing in the
$x$ direction and defocusing in the $y$ direction. The magnetic force leads to
the same refractive powers as presented by Eqs.~\eqref{eq:quad_focus_x} and
\eqref{eq:quad_focus_y}, but with
\begin{equation*}
k_B^2 = \frac{q}{p}\frac{B_{\rm T}}{a}.
\end{equation*}

Quadrupole lenses are typically used as doublets or triplets for
solutions that are focusing in both transverse directions.
Quadrupoles can also be used for transforming asymmetric beams such
as slit beams from a Penning ion source, for example, into a round
beam.

\subsection{Beam emittance}

\begin{figure}[htbp!]
  \centering
  \includegraphics[width=11cm]{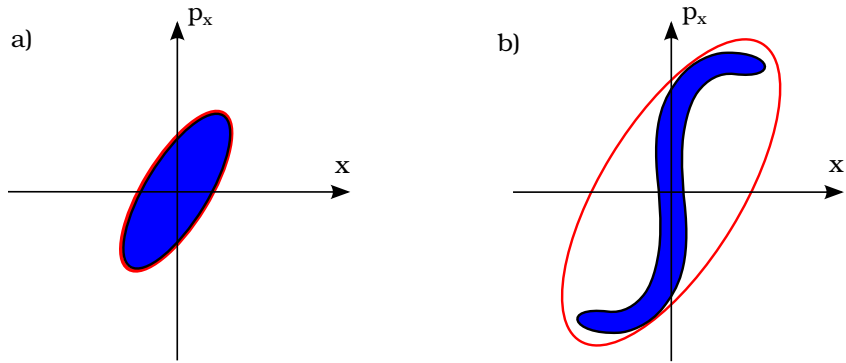}
  \caption{A two-dimensional projection of an ensemble of particles
  (a)~before going through a nonlinear optical system and (b)~after it.
  The area of the particle distribution (shown in blue) is
  conserved but the area of the elliptical envelope (shown in red)
  increases.}
  \label{fig:emittance_conservation}
\end{figure}

Traditionally the emittance is defined as the six-dimensional volume
limited by a contour of constant particle density in the
$(x,p_x,y,p_y,z,p_z)$ phase space. This volume obeys the Liouville
theorem and is constant in conservative fields. With practical
accelerators, a more important beam quality measure is the volume of
the \textit{envelope} surrounding the beam bunch. This is not conserved
generally -- only in the case where the forces acting on the particles
are linear (see Fig.~\ref{fig:emittance_conservation}). Typically in
the case of continuous (or long pulse) beams, where the longitudinal
direction of the beam is not of interest, transverse distributions
$(x,x')$ and $(y,y')$ are used instead of the full phase-space
distribution for simplicity. Also for these distributions the envelope
surrounding the distribution changes when nonlinear forces
(non-idealities of beamline elements, for example) act on the
particles. The size and shape of the transverse distribution envelope
are important quality measures for beams because most complex ion
optical devices such as accelerators have an acceptance window in the
phase space within which they can operate.

\subsubsection{Emittance ellipse}

For calculation and modelling purposes, a simple shape is needed to
model the ion beam envelope in $(x,x')$ phase space. Real well-behaved
ion beams usually have Gaussian distributions in both $x$ and $x'$
directions. Because the contours of two-dimensional (2D) Gaussian distributions
are ellipses, it is an obvious solution to use the ellipse as the model in
2D phase spaces (and ellipsoids in higher
dimensions). The equation for an origin-centred ellipse is
\begin{equation}
    \gamma x^2 + 2\alpha xx' + \beta x'^2 = \epsilon \text{,}
    \label{eq:ellipse}
\end{equation}
where the scaling
\begin{equation}
    \beta \gamma - \alpha^2 = 1
\end{equation}
is chosen. Here $\epsilon$ is the \textit{two-dimensional transverse
emittance}, and $\alpha$, $\beta$ and $\gamma$ are known as the
\textit{Twiss parameters} defining the ellipse orientation and aspect
ratio. The area of the ellipse is
\begin{equation}
    A = \rmpi \epsilon = \rmpi R_1 R_2 \text{,}
\end{equation}
where $R_1$ and $R_2$ are the major and minor radii of the
ellipse. Because of the connection between the area of the ellipse and
$\epsilon$, there is sometimes confusion about whether to include
$\rmpi$ in the above formula for quoted emittance values. The unit of
emittance is often written as $\rmpi${\u}mm{\u}mrad. This is done to
emphasize that the quoted emittance number is the product of the radii
and \textit{not} the area of the ellipse. Always, when communicating
about emittance numbers, it should be clearly indicated what the number
is to avoid confusion.

\begin{figure}[htb]
    \linespread{1.0}
    \centering
    \includegraphics[width=12cm]{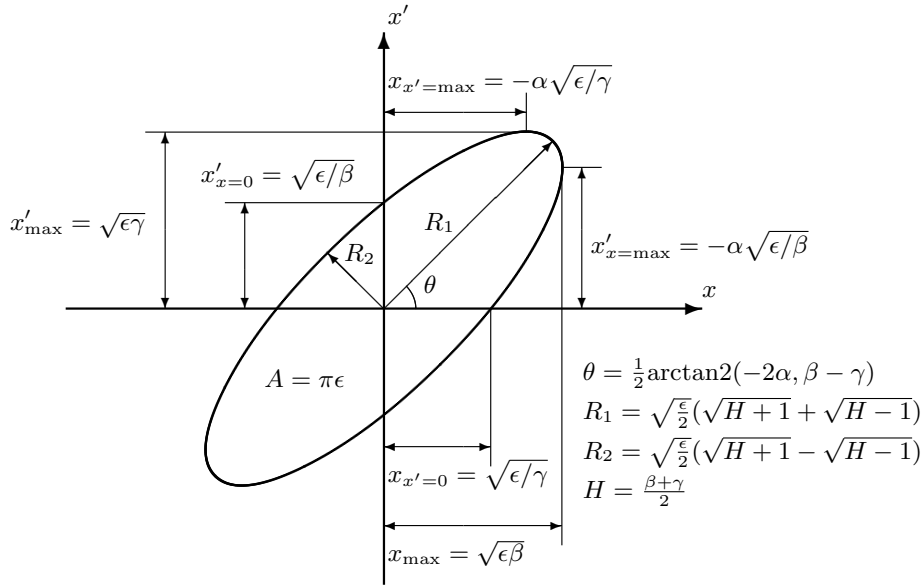}
    \caption{Emittance ellipse geometry with the most important dimensions}
    \label{fig:ellipse_geom}
\end{figure}

From Eq.~\eqref{eq:ellipse} the dimensions of the ellipse can be
calculated. Some of the most important dimensions needed in calculations are
shown in Fig.~\ref{fig:ellipse_geom}.

\subsubsection{Calculating r.m.s.\ emittance}

How do $\epsilon$ and the Twiss parameters relate to phase-space
distributions? How is the envelope defined? There are numerous ways to
fit an ellipse to particle data. Often, a minimum-area ellipse
containing some fraction of the beam is wanted
(e.g.\ $\epsilon_\text{90\%}$), but unfortunately this is difficult to
produce in a robust way. A well-defined way to produce the ellipse
is by using a statistical definition known as the \textit{r.m.s.\ emittance},
\begin{equation}
  \epsilon_{\text{rms}} = \sqrt{\av{x'^2} \av{x^2} - \av{xx'}^2}\text{,}
  \label{eq:e_rms}
\end{equation}
with the expectation values defined as
\begin{eqnarray}
 \av{x^2} &=& \frac{\iint x^2 I(x,x')\,\rmd x\,\rmd x'}{\iint I(x,x')\,\rmd x\,\rmd x'}
 \label{eq:e_rms_exp1} , \\
 \av{x'^2}&=&\frac{\iint x'^2 I(x,x')\,\rmd x\,\rmd x'}{\iint I(x,x')\,\rmd x\,\rmd x'}
 , \\
 \av{xx'}&=&\frac{\iint xx' I(x,x')\,\rmd x\,\rmd x'}{\iint I(x,x')\,\rmd x\,\rmd x'}
 \label{eq:e_rms_exp3} ,
\end{eqnarray}
where $I(x,x')\,\rmd x\,\rmd x'$ is the magnitude of the beam current at the
differential area $\rmd x\,\rmd x'$ of phase space at $(x,x')$. Similarly, the
Twiss parameters can be calculated from the particle distribution with
\begin{equation}
    \alpha = -\frac{\av{xx'}}{\epsilon}, \quad
    \beta = \frac{\av{x^2}}{\epsilon}\quad\mbox{and}\quad
    \gamma = \frac{\av{x'^2}}{\epsilon}.
    \label{eq:twiss}
\end{equation}
For these formulas, it is assumed that the emittance distribution is
centred at the origin, so that $\av{x} = 0$ and $\av{x'} = 0$. With
measured emittances, additional difficulties arise from background
noise and amplifier offsets in $I(x,x')$ data. Filtering methods for
processing experimental data exist, from simple thresholding to more
refined algorithms such as SCUBEEx \cite{SCUBEEX}.

\subsubsection{Amount of beam inside the emittance ellipse}

The meaning of the r.m.s.\ emittance can be more easily understood by
looking at the amount of beam that is enclosed by the ellipse. This, of
course, depends on the particle distribution shape. For real measured
distributions, there is no direct rule. For theoretical known
distributions, this can be calculated. The two most used model
distributions used for beams are the bi-Gaussian and the
Kapchinskij--Vladimirskij (KV) distribution.

The bi-Gaussian distribution oriented along the axes is given by
\begin{equation}
  I(x,x') = \frac{1}{2\rmpi \sigma_x \sigma_{x'}} \exp\left[-\frac{1}{2}
  \left(
  \frac{x^2}{\sigma_x^2} + \frac{x'^2}{\sigma_{x'}^2}
  \right)\right],
\end{equation}
where $\sigma_x$ and $\sigma_{x'}$ are the standard deviations of the
distribution in the $x$ and $x'$ directions. In practice, the distribution
can be additionally rotated by angle $\theta$.

The KV distribution has a uniform beam density inside an elliptical phase space given by
\begin{equation}
  I(x,x') =
  \begin{cases}
    \displaystyle\frac{1}{\rmpi \epsilon} &
    \text{if}\ \gamma x^2 + 2\alpha xx' + \beta x'^2 \le \epsilon , \\[6pt]
    0 &           \text{otherwise}.
  \end{cases}
\end{equation}

\begin{figure}[htb]
    \linespread{1.0}
    \centering
    \includegraphics[width=9.5cm]{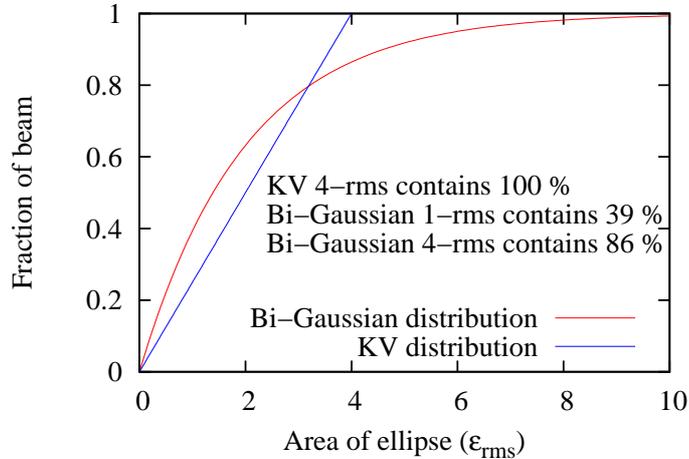}
    \caption{Fraction of beam inside ellipse with $\epsilon =
      q\epsilon_{\text{rms}}$ for hard-edged KV distribution (blue) and
      for Gaussian distribution (red).}
    \label{fig:rms_integral}
\end{figure}

In Fig.~\ref{fig:rms_integral} the beam fraction inside the ellipse
is shown as a function of the area of the ellipse for hard-edged KV
and bi-Gaussian distributions. Because four times the r.m.s.\ ellipse fully
encloses the KV distribution, the so-called \mbox{4-r.m.s.} emittance is often
used as the quoted number instead of the \mbox{1-r.m.s.} emittance.

\subsubsection{Normalization of emittance}

The transverse emittance defined in $(x,x')$ space has the property
that it is also dependent on the longitudinal beam velocity. If the beam
is accelerated and $p_z$ increases, then $x'=p_x/p_z$ decreases. This
effect is eliminated by normalizing the velocity to the speed of light
$c$, which gives
\begin{equation}
  x'_{\rm n} = \frac{p_x}{p_{z1}} \frac{v_{z1}}{c} = \frac{v_x}{c}
  = \frac{p_x}{p_{z2}} \frac{v_{z2}}{c}
\end{equation}
at non-relativistic velocities. The normalized emittance can therefore
be calculated from the unnormalized emittance using
\begin{equation}
  \epsilon_{\rm n} = \epsilon \frac{v_z}{c}.
\end{equation}

\subsubsection{Emittance from plasma temperature}

An ion beam formed by letting charged particles from a plasma be
emitted from a round aperture has an emittance defined by the plasma
ion temperature $T$ and the aperture radius $r$, assuming that the
acceleration to velocity $v_z$ does not add aberrations. This minimum
emittance can be calculated by using
Eqs.~\eqref{eq:e_rms}--\eqref{eq:e_rms_exp3} and using a particle
distribution defined by a circular extraction hole and Gaussian
transverse ion distribution, i.e.
\begin{equation}
  I(x,x') = \frac{2}{\rmpi r^2} \sqrt{r^2-x^2} \, \, \,
  \sqrt{\frac{m}{2\rmpi kT}} \exp \left( \frac{-m(x' v_z)^2}{2kT} \right).
\end{equation}
After normalization, the resulting r.m.s.\ emittance becomes
\begin{equation}
  \epsilon_{\text{rms,n}} = \frac{1}{2} \sqrt{\frac{kT}{m}} \, \frac{r}{c}.
\end{equation}
The calculation can be performed similarly for a slit-beam extraction, which gives
\begin{equation}
  \epsilon_{\text{rms,n}} = \frac{1}{2} \sqrt{\frac{kT}{3m}} \, \frac{w}{c}.
\end{equation}
In the round aperture case, the emittance of the beam is linearly
proportional to the plasma aperture radius. On the other hand, the beam
current is roughly proportional to the area of the plasma
aperture. Scaling of the aperture size does not therefore change the
\textit{beam brightness},
\begin{equation}
  B = \frac{I}{\epsilon_{{\rm n},x} \, \epsilon_{{\rm n},y}}
\end{equation}
in a first approximation.

\subsubsection{Emittance from solenoidal $B$ field}

In electron cyclotron resonance (ECR) and microwave ion sources,
there is a strong solenoidal magnetic
field at the plasma electrode location, where the beam formation
happens. This has a strong influence on the beam quality. As the
particles exit the solenoidal magnetic field, they receive an azimuthal
thrust described by Eq.~\eqref{eq_sol_azimuthal_thurst}. The emittance
of the beam can be calculated outside the solenoid by considering the
particle coordinates far away, where the azimuthal particle motion has
completely changed to radial motion,
\begin{equation}
  r' = \frac{v_r}{v_z} = \frac{v_\theta}{v_z} = \frac{qBr_0}{2mv_z}.
\end{equation}
The r.m.s.\ emittance of the beam can be calculated from the radius of the
constant-current-density beam at extraction and the asymptotic radial
angle,
\begin{equation}
  \epsilon_{\text{rms}} = \frac{1}{4} r_0 r' = \frac{qBr_0^2}{8mv_z}.
\end{equation}
The normalized emittance is given by
\begin{equation}
  \epsilon_{\text{rms,n}} = \frac{qBr_0^2}{8mc}.
  \label{eq:ecr_emittance}
\end{equation}

\begin{figure}[htb]
    \linespread{1.0}
    \centering
    \includegraphics[width=8cm]{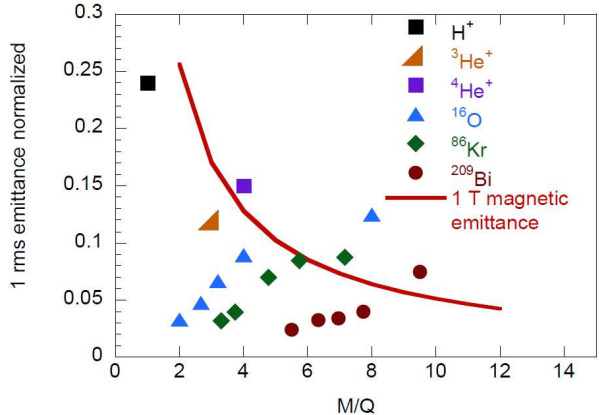}
    \caption{Emittance measurements on the AECR-U ion source for
      various species in comparison with emittances calculated from
      Eq.~\eqref{eq:ecr_emittance} for a magnetic field of $1$~T.
      Reproduced from Ref.~\cite{Leitner2010}.}
    \label{fig:ecr_emittance}
\end{figure}

For ECR ion sources, the effect of the magnetic field dominates
the emittance compared to the effect of the ion temperature, as a result of
the high magnetic fields in these devices. Unfortunately, the formula given
here is not able to predict the emittance values, as measurements
indicate that, for a given element, the higher-charge-state ions have
lower emittances than lower-charge-state ions
(see Fig.~\ref{fig:ecr_emittance}). This trend contradicts
the prediction of Eq.~\eqref{eq:ecr_emittance}. The only possible
interpretation is that the ions are not being extracted from a uniform
plasma. The higher charge states are confined closer to the axis at
the extraction aperture, and therefore their emittance in the beam is
lower. This example shows the additional difficulty in analysing ECR
extractions, as there are no simple self-consistent plasma models
describing the starting conditions for the ions \cite{Leitner2010}.

\subsection{Space charge}

The ion beam charge density
\begin{equation}
  \rho = \frac{J}{v} = \frac{I}{Av}
\end{equation}
plays a major role in beam extraction systems, where current densities
are high and velocities are low compared to other parts of accelerator
systems. The space charge induces forces, which increase the
divergence and emittance, `blowing up' the beam. At higher-energy
parts of the accelerator, the magnetic force generated by the beam
particles starts to compensate the blow-up, but it is insignificant at
$v\ll c$.

\subsubsection{Space-charge effects on beam}
\label{sec:space_charge_effects}

\begin{figure}[htb]
  \linespread{1.0}
  \centering
  \includegraphics[width=10cm]{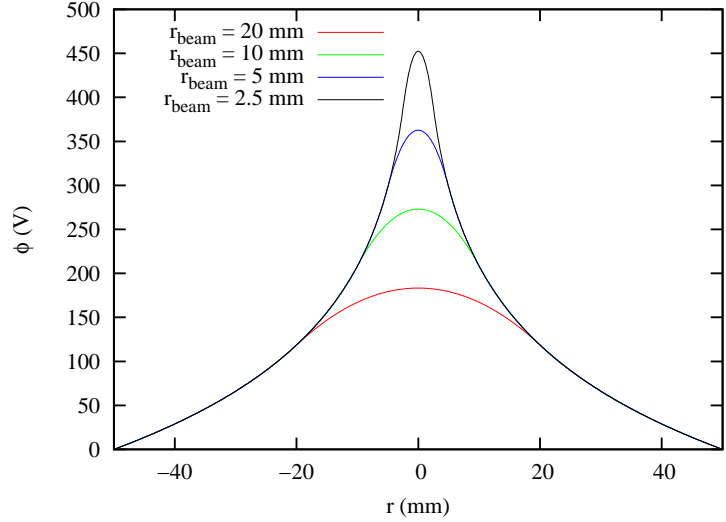}
  \caption{The potential distribution inside a cylindrical 100{\u}mm tube
    with 10{\u}mA, 10{\u}keV proton beam}
  \label{fig:beam_potential}
\end{figure}

Assuming a cylindrical constant-current-density beam with radius $r$
propagating with constant velocity $v_z$, the beam-generated electric
field is given by Gauss's law,
\begin{equation}
  E =
  \begin{cases}
    \displaystyle\frac{I}{2 \rmpi \epsilon_0 v_z} \displaystyle\frac{r}{r_{\text{beam}}^2} &
    \text{if}\ r \le r_{\text{beam}} ,     \\[15pt]
    \displaystyle\frac{I}{2 \rmpi \epsilon_0 v_z} \displaystyle\frac{1}{r} &
    \text{otherwise}.
  \end{cases}
  \label{eq:e_field}
\end{equation}
The potential inside a beam tube with radius $r_{\text{tube}}$ is
therefore
\begin{equation}
  \phi =
  \begin{cases}
    \displaystyle\frac{I}{2 \rmpi \epsilon_0 v}
    \left[ \displaystyle\frac{r^2}{2 r_{\text{beam}}^2} +
    \log\left(\displaystyle\frac{r_{\text{beam}}}{r_{\text{tube}}}\right) -
    \frac{1}{2} \right] & \text{if}\ r \le r_{\text{beam}} ,   \\[15pt]
    \displaystyle\frac{I}{2 \rmpi \epsilon_0 v}
    \log\left(\displaystyle\frac{r}{r_{\text{tube}}}\right) &
    \text{otherwise}.
  \end{cases}
\end{equation}
The potential distribution is plotted in Fig.~\ref{fig:beam_potential} for a 10{\u}mA, 10{\u}keV proton beam inside a 100{\u}mm beam tube.

The electric field in the constant-current-density case, given by
Eq.~\eqref{eq:e_field}, is linear with radius and therefore does not cause
emittance growth, but it does cause increasing divergence of the
beam. A particle at the beam boundary experiences a repulsive force
\begin{equation}
  F_r = qE_r = ma_r = \frac{qI}{2 \rmpi \epsilon_0 r v_z}.
  \label{eq:beam_force}
\end{equation}
Therefore, the particle acceleration is
\begin{equation}
  a_r = \frac{\rmd^2 r}{\rmd t^2} = \frac{\rmd^2 r}{\rmd z^2} \frac{\rmd^2 z}{\rmd t^2}
  = v_z^2 \frac{\rmd^2 r}{\rmd z^2}.
\end{equation}
The particle trajectory is given by the differential equation
\begin{equation}
  \frac{\rmd^2 r}{\rmd z^2} = \frac{1}{v_z^2} a_r = K \frac{1}{r} ,
\end{equation}
where
\begin{equation}
K = \frac{qI}{2 \rmpi \epsilon_0 m v_z^3},
\end{equation}
assuming that the beam divergence is small (i.e.\ Eq.~\eqref{eq:beam_force}
is valid).  The differential equation can be integrated after a change
of variable $\lambda={\rmd r}/{\rmd z}$, which gives
\begin{equation}
  \frac{\rmd r}{\rmd z} = \sqrt{2K \log(r/r_0)},
\end{equation}
assuming ${\rmd r}/{\rmd z}=0$ at $z=0$. The solution is separable and can
be integrated again to achieve the final solution \cite{Humphries1990}
\begin{equation}
  z = \frac{r_0}{\sqrt{2K}} \, F\left(\frac{r}{r_0}\right)
\end{equation}
with
\begin{equation}
  F\left(\frac{r}{r_0}\right) = \int_{y=1}^{r/r_0} \frac{\rmd y}{\sqrt{\log y}}.
\end{equation}
The last integral is not analytic, but can be numerically integrated
for estimates of divergence. As an example, a parallel zero-emittance
beam of $^{181}$Ta$^{20+}$ accelerated with 60{\u}kV has initial radius
of $r_0 = 15${\u}mm. The size of a 120{\u}mA beam after a drift of 100{\u}mm
can be solved from $F(r/r_0)=1.189$, which gives $r = 20${\u}mm.

With practical drifting low-energy beams, a more realistic model for
the beam distribution is bi-Gaussian, for example. This kind of
distribution leads to nonlinear space-charge forces, which cause
emittance growth in addition to increase of beam divergence. Computer
simulations are required to estimate these effects.

\subsubsection{Space-charge compensation}

The potential well of the beam formed by the accelerated charged
particles acts as a trap for oppositely charged particles in areas where
there are no external electric fields to drain the created charges. The
trapped particles compensate the charge density of the beam, decreasing
the depth of the potential well and therefore also decreasing the
magnitude of the beam space-charge effects described above. This
process is called \textit{space-charge compensation}. The most abundant
process for the production of compensating particles is the ionization
of the background gas within the beam. In the case of positive ion
beams, the electrons produced in the background gas ionization are
trapped in the beam, while slow positive ions are repelled to the beamline walls. In the case of negative ion beams, the compensating particles
are the positive ions created in the gas. The creation rate of the
compensating particles can be estimated with
\begin{equation}
  \frac{\rmd n_{\rm c}}{\rmd t} = \Phi n_\text{gas} \sigma_\rmi,
\end{equation}
where $\Phi$ is the flux of beam particles, $n_\text{gas}$ is the gas
density and $\sigma_\rmi$ is the ionization cross-section. If the
creation rate is high enough, the space-charge compensation is finally
limited by the leakage of compensating particles from the potential
well as the compensation factor approaches $100$\%. The compensation
factor achieved in a real system is difficult to estimate accurately
because it depends on the lifetime of the compensating particles in
the potential well. The most important processes affecting the lifetime are
(i)~leakage of particles at the beamline ends, which can be
limited with accelerating einzel lenses or magnetic fields, for
example, (ii)~recombinative processes and (iii)~scattering processes
leading to ejection of particles from the potential well. Assuming
that the creation rate of compensating particles is high, the
time-scale for achieving full compensation is
\begin{equation}
  \tau = \frac{\rho_\text{beam}}{e \, {\rmd n_{\rm c}}/{\rmd t}}
       = \frac{Q}{vn_\text{gas}\sigma_\rmi},
\end{equation}
where $Q$ is the charge state of the beam and $v$ is the velocity of the
beam.  This equation can be used, for example, to estimate if
compensation is possible in pulsed beams.

\begin{figure}[htb]
  \linespread{1.0}
  \centering
  \includegraphics[width=9.0cm]{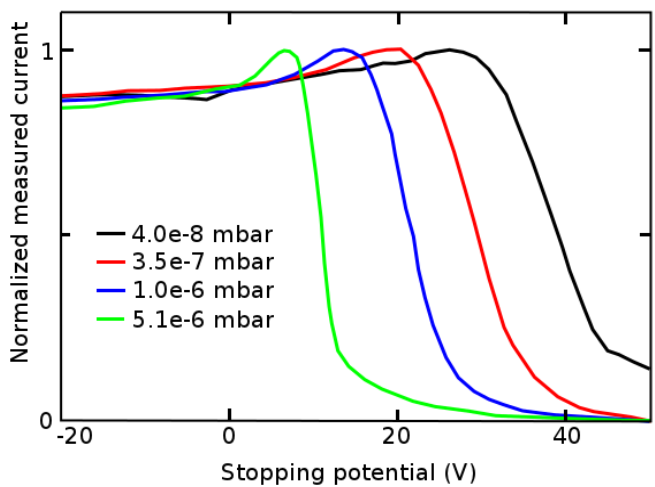}
  \caption{Measurement of ion energy distribution of the ejected ions
    created in the positive ion beam potential well with different gas
    pressures. The potential well depth changes from $\sim$40{\u}V to
    $\sim$10{\u}V when the pressure is increased from $4.0\times10^{-8}${\u}mbar
    to $5.1\times10^{-6}${\u}mbar. Reproduced from Ref.~\cite{Todd2008}.}
  \label{fig:ion_energy_dist}
\end{figure}

In high-beam-intensity LEBT systems, a controlled amount of background
gas is often added to the vacuum chamber to increase the amount of
compensation. A $1$--$2$\% beam loss in the increased ionization
processes can lead to $10$\% increase in total beam transmission due
to decreased divergence.  The magnitude of the compensation can be
estimated, for example, by measurement of the energy distribution of
the ions ejected from the beam as a function of background gas
pressure, as shown in Fig.~\ref{fig:ion_energy_dist}. There are
also particle-in-cell (PIC) computer codes, such as \textsc{Warp}
\cite{WARP} and \textsc{SolMaxP} for modelling the relevant processes
affecting beam compensation. These programs can be used for
analysing beam transport with self-consistent compensation. With other
beam transport programs, a typical solution is to scale the beam
current locally with the space-charge compensation factor estimated by
the user. For more information on space-charge compensation, please
see the chapter by N.~Chauvin in these proceedings dedicated to this topic
\cite{Chauvin2012}.

\section{Beam formation}

In the introduction of this chapter it was assumed that the ion beam is
simply formed by accelerating the plasma particles hitting the plasma
electrode aperture, which separates the quasi-neutral plasma and the
unneutralized beam. In this section, the physics of beam formation
is analysed in more detail.

\subsection{Space-charge-limited emission}

\begin{figure}[htb]
  \linespread{1.0}
  \centering
  \includegraphics[width=0.48\textwidth]{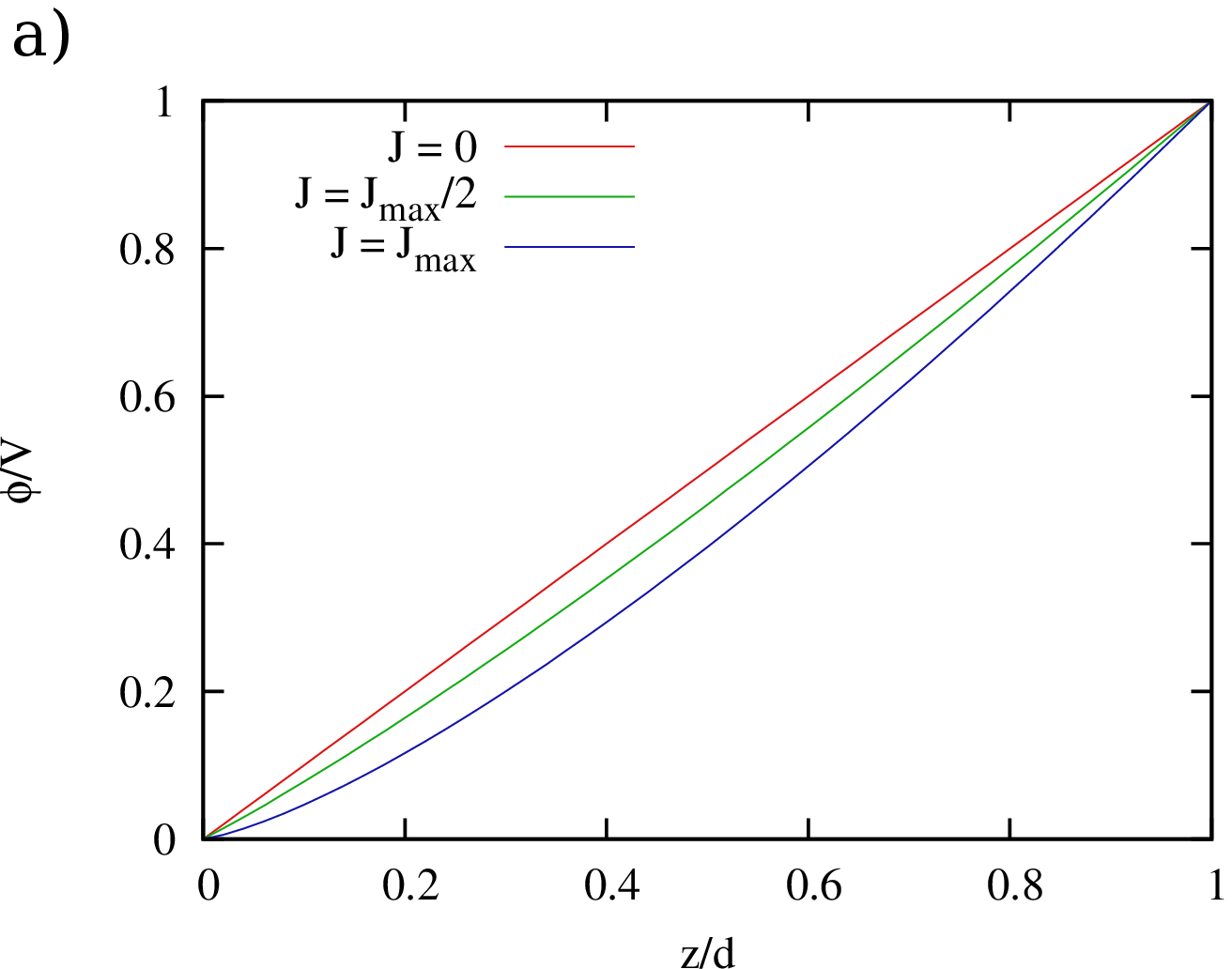}
  \hspace*{0.3cm}
  \includegraphics[width=0.48\textwidth]{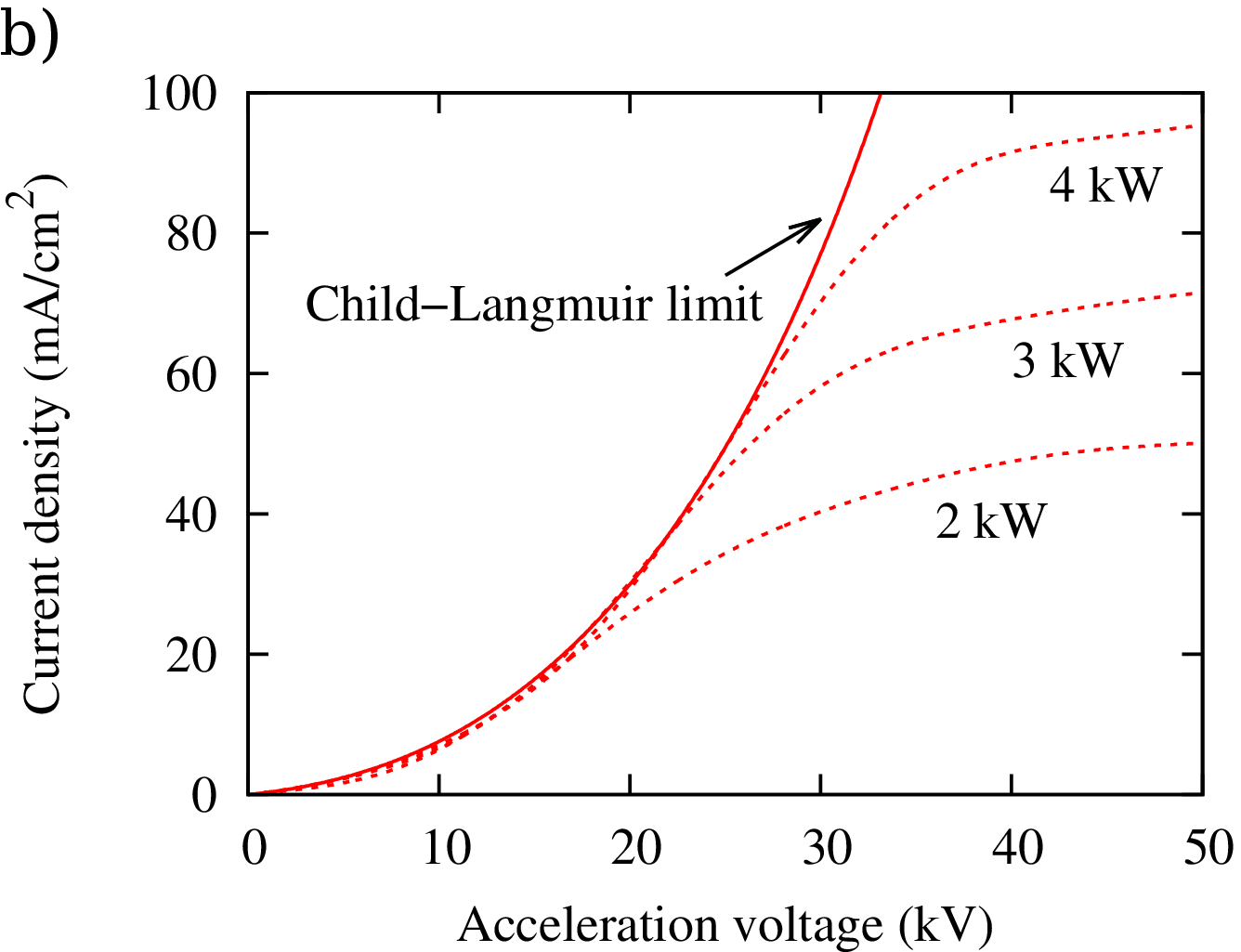}
  \caption{(a) Potential distribution between emission and extractor
    surfaces with different beam current densities in the system. 
    (b)~Typical current--voltage characteristic of plasma extraction. With
    low acceleration voltages, the emission is operating in the
    Child--Langmuir limit. At higher voltages the extracted current
    density saturates due to the emission limit of the plasma (which is
    varied by adjusting the driver power from 2{\u}kW to 4{\u}kW).}
  \label{fig:child}
\end{figure}

In the first acceleration gap, where the beam is formed, the space-charge forces acting on the beam are largest. The situation can be
evaluated in one dimension by assuming a beam starting with zero velocity
with Poisson equation
\begin{equation}
  \frac{\rmd^2 \phi}{\rmd z^2} = -\frac{\rho}{\epsilon_0}
  = -\frac{J}{\epsilon_0} \sqrt{\frac{m}{2q\phi}},
  \label{eq:child_poisson}
\end{equation}
where $z$ is the location, $\phi$ is the gap potential, $J$ is the
beam current density and $\epsilon_0$ is the vacuum permittivity. The
emission surface is at $\phi(z=0)=0$ and the extractor surface is at
$\phi(z=d)=V$. For $J=0$, the potential distribution between the
surfaces is linear. As the emission current density increases, the
electric field at the emission surface decreases until it becomes zero,
as shown in Fig.~\ref{fig:child}(a). At that point the emission
current is at the maximum level, for which Eq.~\eqref{eq:child_poisson} can be solved with the boundary condition
$\frac{\rmd \phi}{\rmd z}(z=0) = 0$. This condition is known as 
space-charge-limited emission, and the resulting limit for the maximum emission
current density can be calculated using the following equation, which
is known as the Child--Langmuir law \cite{Child1911}:
\begin{equation}
  J_\text{max} = \frac{4}{9} \epsilon_0 \sqrt{\frac{2q}{m}} \, \frac{V^{3/2}}{d^2}.
  \label{eq:child}
\end{equation}

Thermionic electron guns are often operated in the space-charge-limited mode for stability and uniformity reasons -- even if the
local emission conditions change, the emitted current stays
constant as long as the emission is limited by the Child--Langmuir
law.

The plasma ion sources are typically operated in emission-limited
mode, i.e.\ the potential difference between the plasma electrode and
puller electrode is made sufficiently large to handle the beam space
charge.  The law in the form shown here is not strictly valid for ion
source plasma extraction because of the effects of plasma
neutralization and higher `starting' velocity of particles in plasma
extraction. The physics of the space-charge limit is still valid and
the Child--Langmuir law \eqref{eq:child} can be used to estimate it.

In any system, the maximum extractable current is dependent on the
geometry, the emission current density and the voltage via the space-charge limit. In the space-charge-limited region, the current is
proportional to $V^{3/2}$. This leads to the definition of the
beam \textit{perveance} as
\begin{equation}
  P = \frac{I}{V^{3/2}},
  \label{eq:perveance}
\end{equation}
which is the proportionality constant describing the system. As long
as the emission is space-charge-limited, the beam perveance is roughly
constant. When the voltage is further increased and the beam emission is
no longer space-charge-limited, the beam perveance decreases. See
Fig.~\ref{fig:child}(b) for an example of the current--voltage
characteristic of a plasma extraction.

\subsection{Electrode geometry}

\begin{figure}[htb]
  \linespread{1.0}
  \centering
  \includegraphics[width=6.00cm]{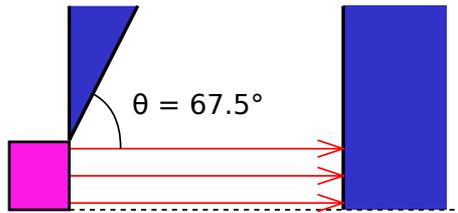}
  \caption{Perfectly parallel extraction of space-charge limited
    surface emission electrons using the Pierce geometry}
  \label{fig:pierce}
\end{figure}

The space-charge forces try to blow up the beam, as was shown
above. This happens especially in the first acceleration gap because
of the low velocity of the beam. To counteract the space-charge forces in
the transverse direction, the electrodes can be shaped in such a way
that the electric field in the first gap is not only accelerating but
also focusing. In the case of space-charge-limited surface-emitted
electrons, there is a perfect solution providing a parallel electron
beam accelerated from the cathode \cite{Pierce1940}. The solution is
to have a field shaping electrode around the cathode (at cathode
potential) in a $67.5^\circ$ angle with respect to the emitting
surface normal, as shown in Fig.~\ref{fig:pierce}. This geometry is
known as Pierce geometry. For plasma ion sources, there is no such
magic geometry because the ions do not start from a fixed surface, but
from plasma with varying starting conditions.

\subsection{Positive ion plasma extraction}
\label{sec:positive_plasma}

In the case of ion plasma extraction, the beam formation is more
complicated than in the case of surface-emitted electrons described
above. The ions are born in the quasi-neutral plasma and get extracted
into the unneutralized beam. It is obvious that the extraction cannot
be modelled without considering the neutralizing effect of the
plasma. The simplest description of the necessary transition layer or
\textit{plasma sheath} was given by Bohm for an ion--electron plasma
\cite{Bohm1949}. The ions are assumed to arrive from the bulk plasma
into the sheath with velocity $v_0$. The charge density of ions can
be calculated by assuming a quasi-neutral situation $\rho_0 = \rho_\rmi =
\rho_\rme$ at the bulk plasma in the plasma potential $\phi = \phi_{\rm P}$,
where $\phi=\phi_\text{wall} = 0$ is the plasma electrode
potential. Using ion continuity $\rho_0 v_0 = \rho_\rmi v_\rmi$ and energy
conservation $m_\rmi v_\rmi^2/2 = m_\rmi v_0^2/2 - q_\rmi (\phi-\phi_{\rm P})$, the ion
density becomes
\begin{equation}
  \rho_\rmi = \rho_0 \sqrt{ 1 - \frac{2q_\rmi (\phi-\phi_{\rm P})}{m_\rmi v_0^2} }.
  \label{bohm_ion_density}
\end{equation}
The electrons are assumed to be in thermal equilibrium and therefore
they follow the Boltzmann distribution
\begin{equation}
  \rho_\rme = \rho_0 \exp \left( \frac{e(\phi-\phi_{\rm P})}{kT_\rme} \right).
  \label{bohm_electron_density}
\end{equation}
The potential in the sheath is described by the Poisson equation
\begin{equation}
  \frac{\rmd^2 \phi}{\rmd x^2} = - \frac{\rho_0}{\epsilon_0} \left[ \sqrt{ 1 -
  \frac{2q_\rmi (\phi-\phi_{\rm P})}{m_\rmi v_0^2} } -
    \exp \left( \frac{e\phi}{kT_\rme} \right) \right].
  \label{sheath_poisson}
\end{equation}
An important feature can be observed from the equation: the shielding
condition, ${\rmd\phi}/{\rmd x}(x=0)=0$ is only fulfilled when the
space charge is non-negative, i.e.\ $\rho_\rmi \ge \rho_\rme$ for all $\phi \le
\phi_{\rm P}$.  The necessary condition
\begin{equation}
  v_0 \ge v_{\rm B} = \sqrt{ \frac{kT_\rme}{m_\rmi} }
\end{equation}
is known as the \textit{Bohm sheath criterion} and $v_{\rm B}$ as the \textit{Bohm
velocity}. The criterion sets a low-velocity limit for ions arriving
at the sheath edge and in most cases the equation holds with equality
\cite{Chapman80}.  The Poisson equation \eqref{sheath_poisson} is
impossible to solve analytically, and often a numerical approach or
approximations are used even in the presented one-dimensional
case. The computational approach that is used in plasma extraction
simulations in higher dimensions is presented in section~\ref{codesBeamExtractTransport}.

\begin{figure}[htb]
  \linespread{1.0}
  \centering
  \includegraphics[height=5.0cm]{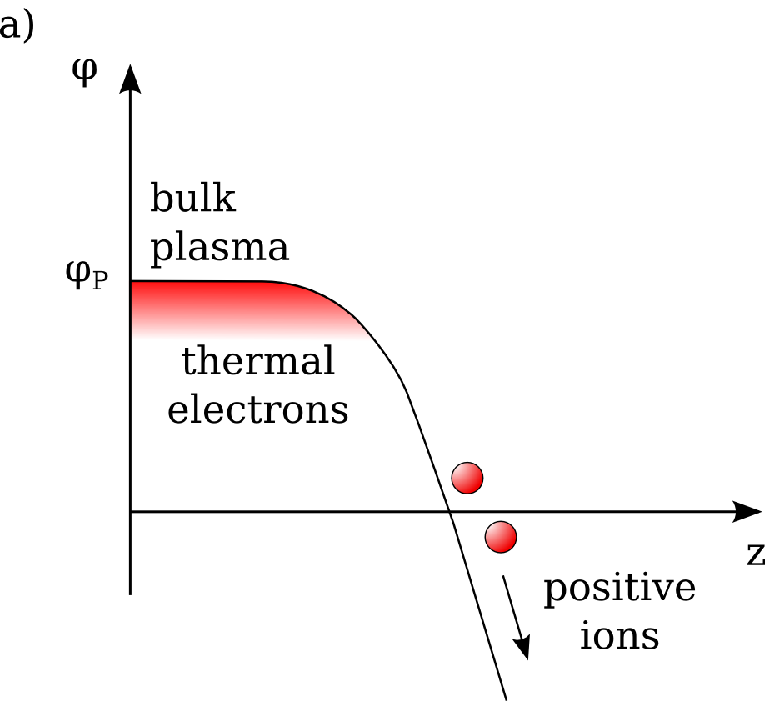}
  \hspace*{1cm}
  \includegraphics[height=6.00cm]{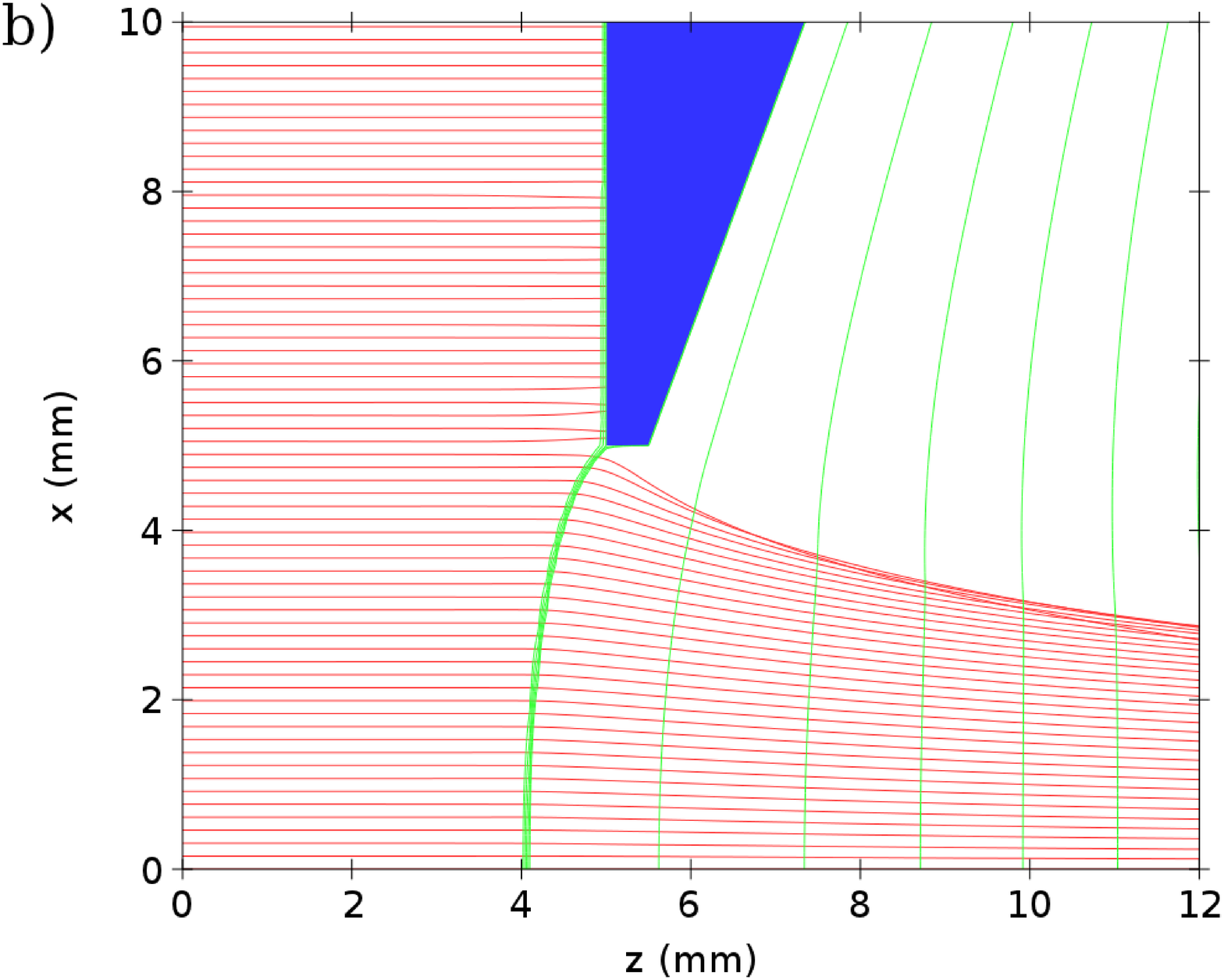}
  \caption{Positive ion extraction. (a)~Schematic representation of the
    model with thermal electrons populating the bulk plasma and
    positive ions accelerated by the extraction field through the
    plasma sheath. (b)~Plot from a 2D simulation of a proton extraction
    with several equipotential lines drawn close to 0{\u}V to visualize
    the location of the plasma meniscus. The ion temperature in the
    simulation is set to zero to show how the particle trajectories
    are accelerated perpendicular to the meniscus after leaving the
    plasma.}
  \label{fig:plasma_model_pos}
\end{figure}

The situation at the plasma extraction is simple according to the
model (see Fig.~\ref{fig:plasma_model_pos}(a)). Positive ions flow
from quasi-neutral bulk plasma into the extraction sheath with velocity
$v_{\rm B}$. The compensating electron density, which is defined by the
potential, is equal to the ion density in bulk plasma and decays
exponentially towards the extraction. Far enough in the extraction,
the compensation becomes (essentially) zero. From the model, it is
obvious that there is no well-defined boundary between neutralized
plasma and the unneutralized extraction. Often, such a boundary would
be useful for judging the focusing action of the electric field close
to the plasma electrode and for communicating about the plasma sheath
shape. Therefore, an equipotential surface at $\phi_\text{wall}$ in the
case of positive ion extraction is often chosen as an artificial
`boundary', known as the \textit{plasma meniscus}. This choice works as
a thought model even though in reality there is no such boundary. See
Fig.~\ref{fig:plasma_model_pos}(b) for a two-dimensional example of
the plasma sheath.

\begin{figure}[htb]
  \linespread{1.0}
  \centering
  \includegraphics[width=0.99\textwidth]{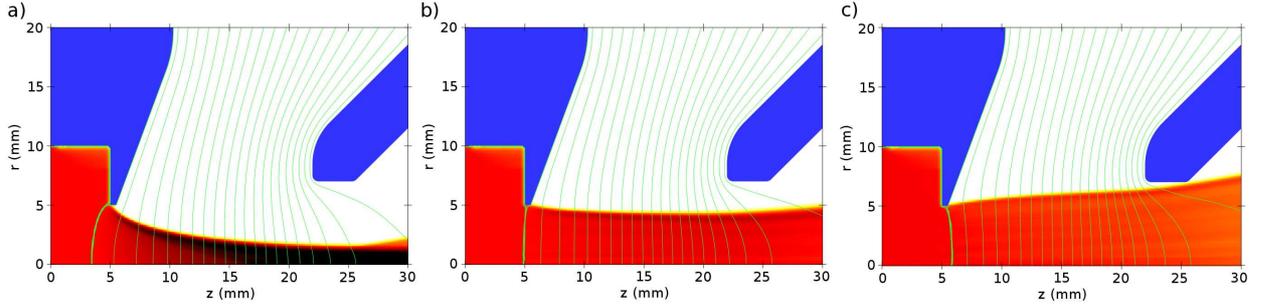}
  \caption{Three simulations of 30{\u}keV proton extraction with
    varying plasma densities. In (a) 25{\u}mA of ions is extracted with
    0.16{\u}mm{\u}mrad r.m.s.\ emittance; in (b) 60{\u}mA with 0.09{\u}mm{\u}mrad;
    and in (c) 95{\u}mA with 0.13{\u}mm{\u}mrad r.m.s.\ emittance. The 
    highest-quality beam is achieved with the flat plasma meniscus.}
  \label{fig:plasma_meniscus}
\end{figure}

The process of beam formation varies not only with the extraction
electric field strength and shape, but also with the properties of the
plasma, i.e.\ plasma density, electron and ion temperatures. In 
Fig.~\ref{fig:plasma_meniscus}, three simulated cases are shown with
differing plasma densities. All other parameters are unchanged. In
case (a) the plasma density is low and 25{\u}mA of protons are
extracted. The strong electric field in the extraction makes the
plasma meniscus concave and the extracted beam is over-focused,
causing increase in beam emittance. In case (b) the plasma density is
higher and 60{\u}mA of beam is extracted. The meniscus shape is almost
flat, which provides the lowest beam emittance. In case (c), the plasma
density is even higher and 95{\u}mA of beam is extracted. The plasma
meniscus is convex, the beam is divergent and the emittance is higher
than in the optimal case. Because of this effect, it is important that
the electric field strength of the extraction system can be somehow
adjusted if changing plasma densities are expected. Possible
adjustments are changing the plasma electrode to puller electrode gap
or changing the puller electrode voltage.

\subsection{Negative ion plasma extraction}
\label{sec:negative_plasma}

\begin{figure}[htb]
  \linespread{1.0}
  \centering
  \includegraphics[height=4.50cm]{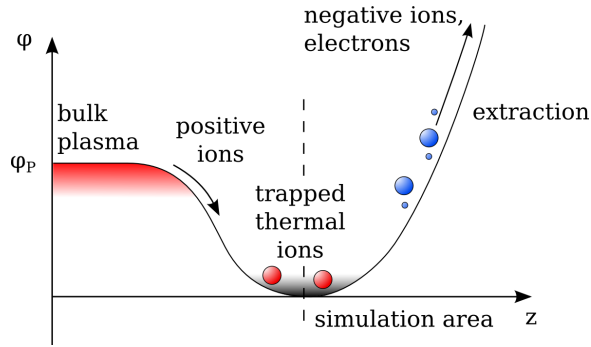}
  \caption{Schematic representation of negative ion plasma extraction
    model with fast positive ions flowing from bulk plasma towards the
    extraction, thermal positive ions populating the potential well at
    the plasma electrode potential, and negative charged particles
    accelerated by the extraction field.}
  \label{fig:plasma_model_neg}
\end{figure}

The negative ion plasma extraction model is similar to the positive ion
extraction model presented above. The bulk plasma is at positive
plasma potential $\phi_{\rm P}$ and it is separated from the plasma
electrode at $\phi=\phi_\text{wall}=0${\u}V by a plasma sheath. It is
assumed that the extractable negative ions, which are either volume- or
surface-produced, are born close to the wall potential and extracted
from a uniform plasma volume. These charges form a potential well and
counteract the formation of a saddle point at the extraction
aperture. The non-existence of the saddle point is supported by the
observed good emittance from H$^-$ ion sources
\cite{Becker04a,Becker04b}. The potential deviates from zero going
into the bulk plasma due to the plasma potential and towards the
extraction due to the acceleration voltage. This potential structure
causes positive ions from the bulk plasma to be accelerated towards
the extraction, having energy $e \phi_{\rm P}$ at the zero potential. These
ions propagate until they are reflected back into the plasma by the
increasing potential in the extraction. The potential well acts as a
trap for thermal positive ions. The negative ions and electrons are
accelerated from the wall potential towards the bulk plasma and more
importantly towards the extraction. A schematic view of the negative ion
extraction model is shown in Fig.~\ref{fig:plasma_model_neg} \cite{Kalvas2011}.

The negative ion plasma sheath from the zero potential towards the
extraction can be described by the Poisson equation
\begin{equation}
  \nabla^2 \phi = - \frac{\rho}{\epsilon_0},
\end{equation}
where $\rho = \rho_{\text{neg}} + \rho_{\text{f}} + \rho_{\text{th}}$.  Here
$\rho_{\text{neg}}$ is the space-charge density of negative particles,
$\rho_{\text{f}}$ is the space-charge density of fast positive ions and
$\rho_{\text{th}}$ is the space-charge density of trapped positive thermal
ions.

The model allows several different negative ion species to be
extracted from the ion source and also many positive ion species to be
used as compensating plasma particles. Each of the thermal ion species
has a separate Maxwellian velocity distribution with the associated
space-charge distribution
\begin{equation}
  \rho_{\textrm{th}} = \rho_{\textrm{th,0}} \exp \left( \frac{-e\phi}{kT_\rmi} \right),
\end{equation}
where $\rho_{\textrm{th,0}}$ is the space-charge density of the
thermal ion species at the wall potential and $T_\rmi$ is the
corresponding thermal ion temperature. The fast ions are decelerated
and turned back into the plasma by the extraction voltage. The space-charge
distribution of the fast ions is defined by the virtual cathode
formation and it is
\begin{equation}
  \rho_{\textrm{f}} = \rho_{\textrm{f,0}} \left( 1 - \frac{e\phi}{E_\rmi} \right),
\end{equation}
at $e\phi < E_\rmi$ and zero otherwise. Here $\rho_{\text{f,0}}$ is the
space-charge density of fast ions at the wall potential and $E_\rmi$ is
the corresponding kinetic energy, which should be around $e\phi_{\rm P}$ as
the particles are flowing from the bulk plasma. The quasi-neutrality
of the plasma requires $\rho_{\text{neg}} + \rho_{\text{f}} +
\rho_{\text{th}} = 0$ at $\phi = 0${\u}V.

In the negative ion extraction, the plasma sheath acts similarly to
the positive ion extraction case. The smallest beam emittance is
achieved with extraction field optimized for the plasma density of the
ion source. The biggest difference from the positive ion case is that,
with a positive puller electrode voltage, also electrons will be
extracted from the plasma in addition to the negative ions. Depending
on the ion source, the amount of co-extracted electrons may be as high
as 100--200 times the amount of negative ions extracted or as low as~1
as is the case in caesiated surface production H$^-$ sources.
Especially in the cases where the amount of electrons is high, the
electrons need to be dumped in a controlled manner as soon as possible
to avoid unnecessary emittance growth due to the additional space
charge. Often the electron beam current is so high that the dumping
cannot be done at the full beam energy required by the application. In
other words, the electron beam has to be dumped on an intermediate
electrode at lower potential than ground.

Typically the electron dumping is done by utilizing a magnetic dipole
field generated with permanent magnets. As the magnetic field also
deflects the negative ion beam, the negative ion source may be mounted
to the rest of the beam transport line at an angle to compensate for
the deflection. Another solution is to use another dipole field
in the opposite direction to correct the angle of the negative ion
beam back perpendicular to the original axis. The resulting offset in
the beam centre has to be corrected using deflector plates, $XY$
magnets or mechanical offset.

\begin{figure}[!htbp]
  \linespread{1.0}
  \centering
  \includegraphics[height=7.0cm]{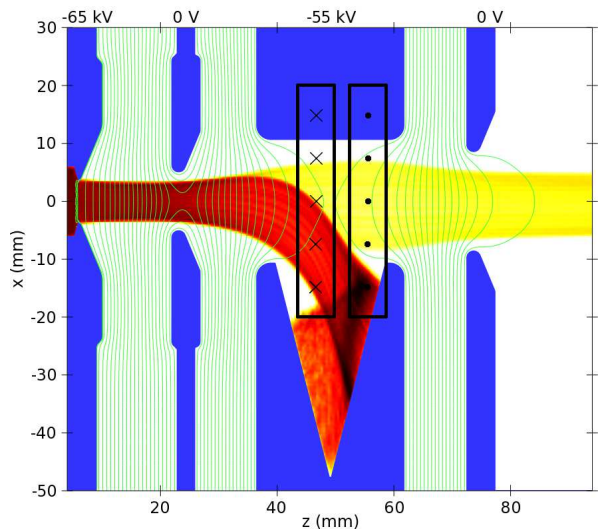}
  \caption{An extraction system designed for the SNS for transporting
    up to 100{\u}mA of H$^-$ and for dumping up to 1{\u}A of electrons into
    a V-shaped electron dump. The puller electrode can be adjusted to
    cope with varying current levels. The electron dump (einzel)
    electrode has permanent magnets embedded for a dipole--antidipole
    field structure \cite{Kalvas2012}.}
  \label{fig:SNS_new}
\end{figure}

There are three locations, which are generally used for dumping the
electron beam in negative ion sources: 1. Puller electrode with low
voltage with respect to the ion source can be used as an electron dump
\cite{Kuo1996,Kalvas2011}, 2. a separate intermediate electrode before
the puller electrode can be used if a higher voltage is needed on the
puller electrode \cite{Keller2002} or 3. the electrons are dumped on
an electrode after the puller electrode, as shown in
Fig.~\ref{fig:SNS_new}. The third method has the advantage that the
voltage of the puller electrode can be optimized for plasma density
matching without affecting the electron dumping
\cite{Midttun2012,Kalvas2012}.

\section{Computer codes for beam extraction and transport problems}
\label{codesBeamExtractTransport}

The modelling and analysis of modern beam extraction and transport
systems is so complicated that most of the work is done using
specialized computer simulation codes. In this section, a brief
overview of the wide spectrum of codes is presented.

\subsection{Transfer matrix codes}

Traditionally, modelling of ion optical transport lines has been done
using the transfer matrix formalism, which describes the effect of ion
optical elements (and drifts between them) on trajectories using
transfer matrices. In the computer program \textsc{Transport}
\cite{Transport}, for example, the properties of a single-particle
trajectory at any given point of the transport line are described by
a vector $\bvm{X}=(x,x',y,y',l,\delta)$, where $x$ and $y$ are the
transverse displacements of the trajectory with respect to the
reference trajectory, $x'$ and $y'$ are the tangents of angles of the
trajectory with respect to the reference trajectory, $l$ is the path-length difference between the trajectory and the reference trajectory,
and $\delta=\Delta p/p$ is the fractional momentum deviation of the
trajectory from the reference trajectory. The trajectory component
$X_i$ after propagation through an ion optical element can be
calculated from
\begin{equation}
  X_i = \sum_j Y_j \left\{ (X_i \mid Y_j) +
    \sum_k \frac{Y_k}{2} \left\{ (X_i \mid Y_jY_k) +
    \sum_l \frac{Y_l}{3} \left\{ (X_i \mid Y_jY_kY_l) +
    \cdots \right\} \right\} \right\},
    \label{eq:matrix_transformation}
\end{equation}
where $Y_i$ are the components of the trajectory before the ion
optical element, and $(X_i \mid Y_j)$, $(X_i \mid Y_jY_k)$, $(X_i \mid
Y_jY_kY_l)$, \ldots\ are the first-order, second-order, third-order,
\ldots\ transfer coefficients. This can be described as matrix--vector
multiplication, with a $6\times 6$ matrix in first order, a $6\times
6^2$ matrix in second order, a $6\times 6^3$ matrix in third order,
etc., using an extended trajectory vector containing also the
higher-order terms, for example, $(x, x', y, y', l, \delta, x^2, xx',
xy, xy', xl, x\delta, x'^2, x'y, \ldots, \delta^2)$ in the second
order. If the element matrices are expanded as square matrices, the
propagation of the trajectory through the whole beamline can be
calculated from the product of the element matrices $\bvm{R}$ instead
of using Eq.~\eqref{eq:matrix_transformation} for each element
$\bvm{R}(1),\ \bvm{R}(2),\ \ldots$. The propagation of trajectory
$\bvm{Y}$ through the whole system is given by
\begin{equation}
  \bvm{X} = \bvm{R} \, \bvm{Y} = \bvm{R}(n) \cdots \bvm{R}(3)\,\bvm{R}(2)\,\bvm{R}(1)\,\bvm{Y}.
\end{equation}
The cumulative matrix $\bvm{R}$ can be used to calculate several
trajectories through the transport line much faster than just by using
the projection from element to element one after another.

The first-order approximation matrices can be easily derived
analytically by assuming that the displacements and angles relative to
the reference trajectory are small enough to justify the truncation of
higher-order terms. For example, the first-order matrix for a
magnetic quadrupole lens is
\begin{equation}
  \bvm{R} = \left(
  \begin{array}{cccccc}
    \cos kL    & (1/k) \sin kL & 0 & 0 & 0 & 0 \\
    -k \sin kL & \cos kL             & 0 & 0 & 0 & 0 \\
    0 & 0 & \cosh kL & (1/k) \sinh kL & 0 & 0 \\
    0 & 0 & k \sinh kL & \cosh kL & 0 & 0 \\
    0 & 0 & 0 & 0 & 1 & {L}/{\gamma^2} \\
    0 & 0 & 0 & 0 & 0 & 1
  \end{array} \right),
\end{equation}
where $L$ is the effective length of the quadrupole,
\begin{equation*}
k_B^2=\frac{B_{\rm T}}{a}\frac{q}{p}
\end{equation*}
and $\gamma$ is the relativistic
factor. Some ion optical effects can also be derived to higher-order
approximations analytically. Usually the highest-order matrices are
constructed using numerical integration of known fields.

In addition to transporting single trajectories or distributions, the
same matrix formalism can be used to transport elliptical beam
envelopes through the transport line in the first order. In this case,
the beam envelope (in one dimension) is described by the beam matrix
\begin{equation}
  \boldsigma = \epsilon \left(
  \begin{array}{cc}
    \beta & -\alpha \\
    \alpha & \gamma
  \end{array}
  \right),
\end{equation}
where $\alpha$, $\beta$ and $\gamma$ are the Twiss parameters and
$\epsilon$ is the beam emittance. The transformation of the envelope
$\boldsigma$ through a system described by matrix $\bvm{R}$ can be calculated with
\begin{equation}
  \boldsigma' = \bvm{R} \, \boldsigma \, \bvm{R}^{\rm T}.
\end{equation}

Programs using the transfer matrix formalism are typically used for
modelling long beam transport lines at relatively high energies, where
the beam size and angles are small. The calculation of trajectories is
fast, which allows automatic optimization of ion optics, etc. Many of
the programs even have models for space-charge-induced divergence
growth (similar to what was presented in section~\ref{sec:space_charge_effects}) 
and/or r.m.s.\ emittance growth modelling
using the paraxial approximation for particle distributions, which
makes them more suitable for modelling the low-energy and high-current
systems such as LEBT and accelerator injection systems. The use of the
space-charge models, of course, makes the calculation slower, but the
codes are still much faster than the ray-tracing codes discussed in
section~\ref{rayTraceExtract}. Some of the most commonly used codes of this type include the following:
\begin{itemize}
  \item \textsc{Cosy Infinity} \cite{Cosy}, up to infinite order, no
    space charge modelling
  \item \textsc{DiMad} \cite{Dimad}, up to third-order calculation,
    space charge of KV beam
  \item \textsc{Gios} \cite{Gios}, up to third-order calculation,
    space charge of KV beam
  \item \textsc{Path Manager (Travel)} \cite{Path}, up to second
    order, more advanced space-charge modelling for particle
    distributions (mesh or Coulomb model)
  \item \textsc{Trace-3D} \cite{Trace}, mainly linear with space
    charge of KV beam
  \item \textsc{Transport} \cite{Transport}, up to third-order
    calculation, no space-charge modelling
\end{itemize}

The transfer matrix codes are unfortunately not usable for
electrostatic extraction systems, because no general matrices exist
for describing the optics of the electrode systems.

\subsection{Ray-tracing and extraction codes}
\label{rayTraceExtract}

\begin{figure}[htb]
  \linespread{1.0}
  \centering
  \includegraphics[width=12.00cm]{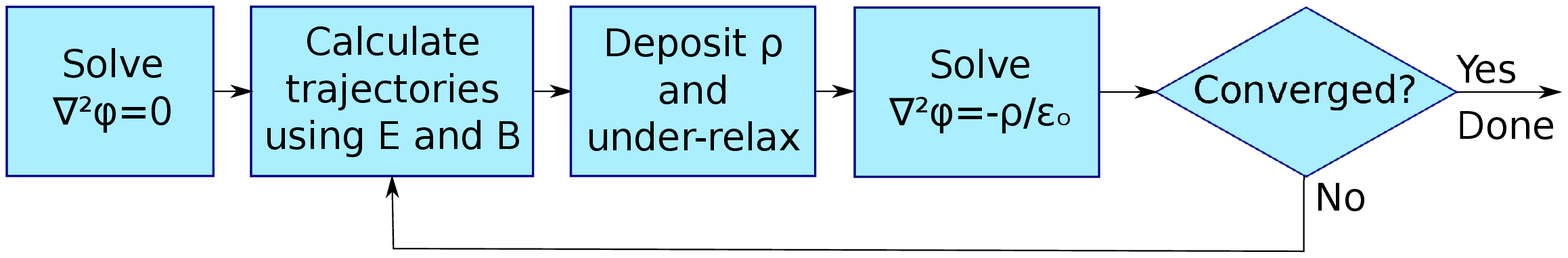}
  \caption{Flow diagram for the iterative solution of high-space-charge 
  beam transport with ray tracing}
  \label{fig:iteration}
\end{figure}

For systems where the approximations made in the transfer matrix codes
are not valid, another more fundamental method has to be used. An
approach taken by so-called \textit{ray-tracing codes} is to directly
integrate the particle equation of motion (Eq.~\eqref{eq:motion})
using mesh-based maps for $\vec{E}$ and $\vec{B}$ fields. In the
extraction system case, it is typically assumed that the
$\vec{B}$ field is defined only by external sources, i.e.\ the 
beam-generated magnetic field is negligible.  On the other hand, the space
charge of the beam plays a major role, and the self-consistent solution
of the beam transport can be found using the iterative approach shown in
Fig.~\ref{fig:iteration}.

The iteration starts with computation of
the electric potential on the system using the Laplace equation. The
electric field is then calculated from the potential, and particles are
traced through the system. The current carried by the particles is
deposited onto the underlying mesh for the following solution of the
Poisson equation for a new estimate of the electric potential. The
iteration is continued like this until convergence is reached. A code
of this type may also have a nonlinear Poisson solver for taking into
account the compensating particles of the plasma using analytical
formulas, as described in sections \ref{sec:positive_plasma} and
\ref{sec:negative_plasma}. Pioneering work in the field using this
method was done by S.~A.~Self \cite{Self1963} and J.~H.~Whealton
\cite{Whealton77,Whealton78} in the 1960s and 1970s. This type
of code is often called an \textit{extraction code} or \textit{ion gun
code}. The most used codes of this type include the following:
\begin{itemize}
  \item \textsc{IGun} \cite{Becker04a}, a code with plasma modelling
    for negative and positive ions, only 2D and cylindrically
    symmetric geometries
  \item \textsc{PbGuns} \cite{PbGuns}, plasma modelling for positive
    and negative ions, only 2D and cylindrically symmetric geometries
  \item \textsc{Kobra} \cite{Kobra}, a more advanced 3D $E$ field
    solver, positive ion plasma modelling, simple particle-in-cell
    capability
  \item \textsc{IBSimu} \cite{Kalvas2010,IBSimu_web}, 1D, 2D or 3D and
    cylindrically symmetric $E$ field solver, plasma modelling for
    positive and negative ions
\end{itemize}

In the following, the typical methods used in extraction codes are
described in more detail. The particular choices presented here are
those that are used in the \textsc{IBSimu} code, which was written
by the author, but many codes of this type use the same or similar
methods.

\subsubsection{Electric potential and field}

The geometry of the simulation domain is discretized using a regular
grid, where the coordinates of calculation nodes can be calculated from
integer indices $i$, $j$, $k$, with $x_i = x_0 + ih$, $y_j = y_0 +
jh$ and $z_k = z_0 + kh$.  The Poisson equation
\begin{equation}
  \nabla^2 \phi =
  \frac{\rmd^2 \phi}{\rmd x^2} + \frac{\rmd^2 \phi}{\rmd y^2} + \frac{\rmd^2 \phi}{\rmd z^2} = -\frac{\rho}{\epsilon_0}
\end{equation}
in three dimensions is discretized using the finite difference method
(FDM), i.e.\ by replacing the derivatives with finite differences. The
Poisson equation becomes
\begin{equation}
  \frac{\phi_{i-1,j,k} + \phi_{i+1,j,k} +
        \phi_{i,j-1,k} + \phi_{i,j+1,k} +
        \phi_{i,j,k-1} +\phi_{i,j,k+1} - 6\phi_{i,j,k}}{h^2}
  = - \frac{\rho_{i,j,k}}{\epsilon_0}
  \label{eq:fdm_poisson}
\end{equation}
for nodes that do not have electrodes as close neighbours. The nodes that
are close to electrodes use a modified form of the Poisson equation with
uneven node distances to take into account the real distance of the
calculation node from the surface for more accurate, smooth solution
near the surface. The finite difference for the partial derivative in the
$x$ direction becomes
\begin{equation}
  \frac{\rmd^2 \phi}{\rmd x^2} =
  \frac{\beta \phi(x_0-\alpha h) - (\alpha + \beta) \phi(x_0) +
    \alpha \phi(x_0+\beta h)}
       {\frac{1}{2} (\alpha+\beta) \alpha \beta h^2},
  \label{eq:fdm_poisson_uneven}
\end{equation}
where $\alpha h$ is the distance from $x_i$ to the location where the
potential is known in the negative $x$ direction and $\beta h$ is the
distance from $x_i$ to the location where the potential is known in the
positive $x$ direction. The near-solid distances ($\alpha$ and $\beta$ coefficients) are stored in a table for all calculation nodes that are neighbours to surfaces.  The nodes that are on the boundaries of
the simulation domain have to be constrained by a boundary condition,
either a Dirichlet boundary condition (fixed potential)
\begin{equation}
  \phi_{i,j,k} = \phi_\text{const.}
  \label{eq:fdm_dirichlet}
\end{equation}
or a Neumann boundary condition (fixed derivative with respect to the
normal of the boundary)
\begin{equation}
  \frac{-3\phi_{i,j,k} + 4\phi_{i+1,j,k} - \phi_{i+2,j,k}}{2h} =
  \left(\frac{\rmd \phi}{\rmd x}\right)_\text{const.},
  \label{eq:fdm_neumann}
\end{equation}
in the case of a boundary with normal in the positive $x$ direction. The
numerical formulation of the Poisson equation then becomes a system of
$N$ simultaneous equations, where $N$ is the number of free
(non-constant) nodes. The problem is often described as a matrix
equation
\begin{equation}
  \bvm{A}\,\boldphi = \bvm{B},
\end{equation}
where $\bvm{A}$ is an $N\times N$ matrix of coefficients, $\bvm{B}$ is a vector
of coefficients from Eqs.~\eqref{eq:fdm_poisson}--\eqref{eq:fdm_neumann} and $\boldphi$ is the solution electric potential vector. In the case of plasma modelling, the
Poisson equation contains an analytical term for the space-charge
density of the compensating plasma particles. This leads to a nonlinear
Poisson problem, which is typically formulated as
\begin{equation}
  \bvm{A}(\boldphi) = 0.
  \label{eq:nonlinear_Poisson}
\end{equation}
In the typical scale of systems being investigated, $N$ is
$10^6$--$10^8$, which makes the solution of Eq.~\eqref{eq:nonlinear_Poisson} computationally intensive. The problem is
typically solved using $N$-dimensional Newton--Raphson methods with
iterative linear solvers.

\begin{figure}[htb]
  \linespread{1.0} \centering
  \includegraphics[width=3.5cm]{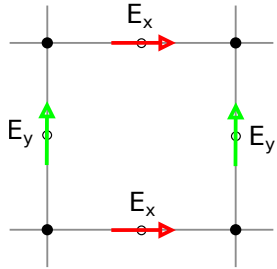}
  \caption{The electric field is evaluated on a different mesh with
    $h/2$ offset from the electric potential mesh to avoid further approximation.}
  \label{fig:efields}
\end{figure}

The electric field is calculated using meshes with $h/2$ location
offset from the electric potential map nodes to avoid making further
approximations. For $E_x$ field the node location is $x_i = x_0 + h/2 +
ih$, while $y_j$ and $z_k$ coordinates of the nodes are as
before. The electric field is evaluated with
\begin{equation}
  E_{x,i,j,k} = \frac{\phi_{i,j,k} - \phi_{i+1,j,k}}{h},
\end{equation}
and similarly for $E_y$ and $E_z$ as shown in Fig.~\ref{fig:efields}. The near-solid distances ($\alpha$ and $\beta$)
described before are also used here to modify the electric field
calculation near the solids.

\subsubsection{Trajectory calculation}

The particles in this type of simulation represent several physical
particles, and therefore they are typically called \textit{macro-particles}.
Each particle is given charge $q$ and mass $m$, which are parameters
of the physical particle, and current $I$, which is a parameter of the
macro-particle. The particle also has location $\vec{x}$ and velocity
$\vec{v}$, which are set according to the definition given by the
user. In a typical case, where particles are starting inside the
plasma, the location is sampled from a quasi-random distribution to
fill a cylindrical surface on the boundary of the simulation
domain. The longitudinal $z$ component of the velocity is defined to be
larger than the Bohm velocity, while the transverse components are
sampled from a Gaussian distribution with standard deviation $\sigma_v
= \sqrt{T_\rmi e/m}$, where $T_\rmi$ is the ion transverse temperature. The
code also contains many other possibilities for beam definition, such
as rectangular beams, beams with KV- or Gaussian-distributed emittance
pattern or definition of beam particle-by-particle.

The propagation of particles in the simulation domain is calculated by
integrating the particle equation of motion in the form of a set of
ordinary differential equations. Using the non-relativistic
approximation in 3D, the set is
\begin{eqnarray}
  \frac{\rmd x}{\rmd t}   &=& v_x ,\\
  \frac{\rmd y}{\rmd t}   &=& v_y ,\\
  \frac{\rmd z}{\rmd t}   &=& v_z ,\\
  \frac{\rmd v_x}{\rmd t} &=& a_x = \frac{q}{m}(E_x + v_y B_z - v_z B_y) ,\\
  \frac{\rmd v_y}{\rmd t} &=& a_y = \frac{q}{m}(E_y + v_z B_x - v_x B_z) ,\\
  \frac{\rmd v_z}{\rmd t} &=& a_z = \frac{q}{m}(E_z + v_x B_y - v_y B_x).
\end{eqnarray}
The integration of the system of equations is done with a Runge--Kutta
Cash--Karp adaptive algorithm and, at each step, the full set of
particle coordinates $(t,x,v_x,y,v_y,z,v_z)$ is stored.

\subsubsection{Space-charge deposition}

\begin{figure}[htb]
  \linespread{1.0}
  \centering
  \includegraphics[height=4.00cm]{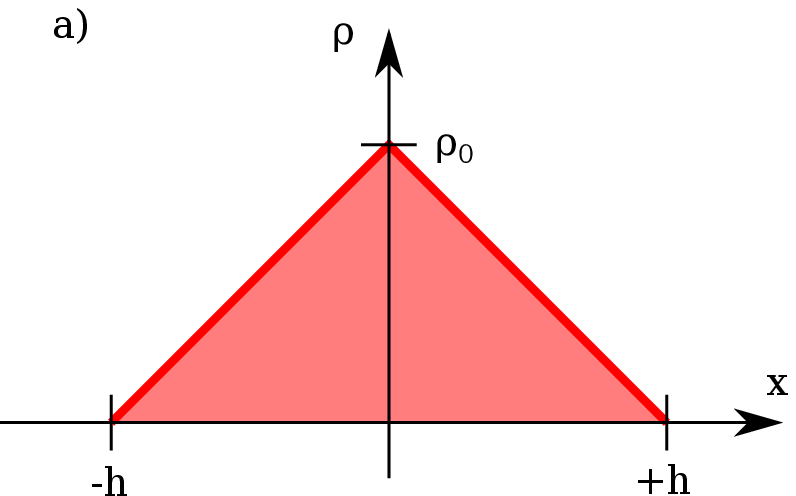}
  \hspace*{1cm}
  \includegraphics[height=4.00cm]{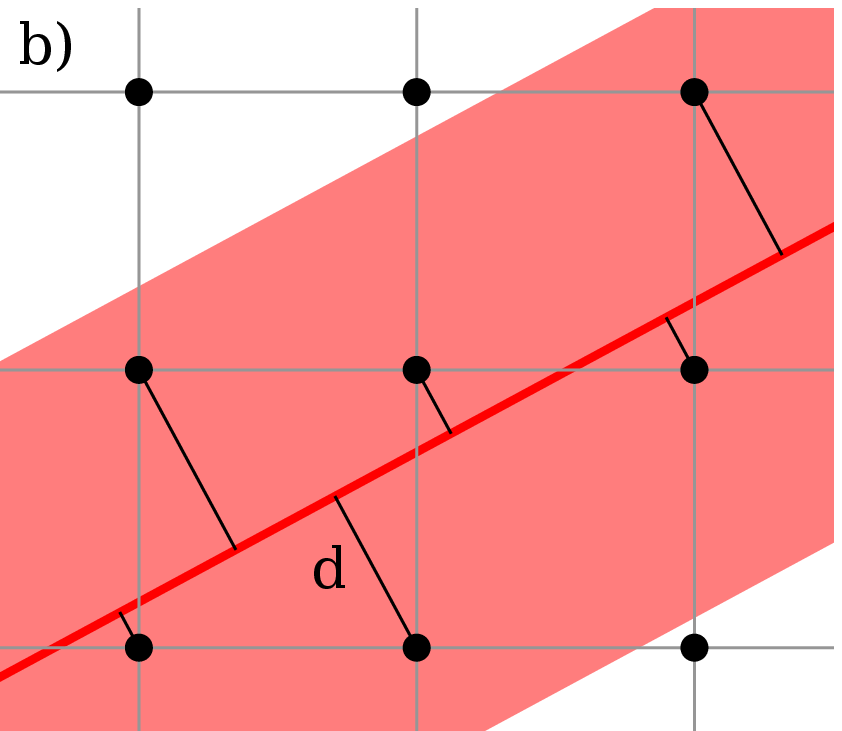}
  \caption{(a) In 2D the particle shape is a triangle with base width
    of $2h$ and height $\rho_0={\lambda}/{hv}$, where $\lambda$ is
    line charge density carried by the 2D trajectory. (b)~As the particle
    passes through the domain, it covers mesh nodes with space-charge
    density according to the particle shape. The space charge deposited
    to mesh nodes is calculated according to the closest distance $d$
    of the node from the trajectory.}
  \label{fig:space_charge}
\end{figure}

The particle trajectories are used to calculate the space-charge
density in the simulation domain. This is done by depositing the
charge carried by the trajectories into the nodes of the mesh in which
the Poisson equation is solved. Each particle trajectory carries
current $I$. The simulated macro-particles must have a finite size
so that space charge can be defined. The simplest scheme for space-charge deposition would be to assume a box-like particle shape with
\begin{equation}
  \rho(x_1,x_2) =
  \begin{cases}
    \displaystyle\frac{I}{h^2v} & \text{if}\ \ {-}\!\frac{1}{2}h < x_1 < \frac{1}{2}h\ \ \text{and}\ \ {-}\!\frac{1}{2}h < x_2 < \frac{1}{2}h ,\\[6pt]
    0            & \text{elsewhere},
  \end{cases}
\end{equation}
where $x_1$ and $x_2$ are the coordinates transverse to the
trajectory. This is known as the closest-node weighting and it is
prone to numerical noise. A better solution is to distribute the space
charge to closest nodes with bilinear weighting. The particle space-charge distribution
\begin{equation}
  \rho(x_1,x_2) =
  \begin{cases}
    \displaystyle\frac{I}{h^2v}(1-|x_1/h|)(1-|x_2/h|) & \text{if}\ \ {-}\!h < x_1 < h\ \ \text{and}\ \ {-}\!h < x_2 < h ,\\[6pt]
    0            & \text{elsewhere},
  \end{cases}
  \label{eq:bilinear_space_charge}
\end{equation}
leads to much smoother space-charge densities. The charge deposition
to the nodes is done based on the closest distance from the trajectory,
as shown in Fig.~\ref{fig:space_charge}.

Even with the smooth space charge deposition of
Eq.~\eqref{eq:bilinear_space_charge}, the overall iteration may not
converge in all cases. In a typical case a region of space gets a high
$\rho$, which leads to electric fields directing the particles away
from the region in the next iteration round. This will lead to a low
$\rho$, which then leads to electric fields directing the particles
into the area. This phenomenon happens easily especially where the
particle velocities are low and trajectories get deflected by low
electric fields. A simple method for suppressing the effect is to use
under-relaxation of the space-charge map to avoid the over-shooting of
$\rho$.

\subsection{Other useful computer codes for ion optics}

Other computer codes that may be useful for ion optics include
\textsc{Poisson Superfish} \cite{Poisson_sf} and \textsc{FEMM}
\cite{FEMM}, which can be used to calculate magnetostatics in 2D and
cylindrically symmetric geometries using the finite element method, and
\textsc{Radia-3D} \cite{Radia}, which uses boundary integral methods
to solve magnetostatic problems in 3D. 
Commercial computational suites such as \textsc{Cobham Vector Fields}
\cite{Cobham}, \textsc{Comsol Multiphysics} \cite{Comsol} and
\textsc{IES Lorentz} \cite{IES} provide advanced finite element- and
boundary element-based field solvers. These packages provide nicely
packaged, easily used tools with graphical user interfaces for typical
problems such as electrostatic and magnetostatic field solution, heat
transfer and even charged-particle tracers. Unfortunately, serious
plasma modelling capabilities are still missing from the commercial
packages. All of the above-mentioned codes can be used together with
\textsc{IBSimu}, for example, for providing $B$ field maps of magnetic
elements.

There also exist many specialized programs for modelling space-charge
compensation, beam bunching, cyclotron injection, radio frequency quadrupole 
(RFQ) acceleration, collisional ion source plasmas, etc., 
with many different methods. The
reader is encouraged to seek more information about such software from
the literature and the World Wide Web.


\begin{thebibliography}{99}

\bibitem{Thomae2002}
R.~Thomae, R.~Gough, R.~Keller, K.~N.~Leung, T.~Schenkel, A.~Aleksandrov, M.\ Stockli and R.\ Welton, Beam measurements on the H$^-$ source and low energy beam transport system for the Spallation Neutron Source. \textit{Rev. Sci.  Instrum.} \textbf{73} (2002) 2016.

\bibitem{Liebl2008} 
H.~Liebl, \textit{Applied Charged Particle Optics} (Springer, Berlin, 2008).

\bibitem{Wollnik1987} 
H.~Wollnik, \textit{Optics of Charged Particles} (Academic Press, Orlando, FL, 1987).

\bibitem{Kumar09} 
V.~Kumar, Understanding the focusing of charged particle beams in a solenoid magnetic field. \textit{Am. J. Phys.} \textbf{77} (2009) 737.

\bibitem{SCUBEEX} 
M.~P.~Stockli, R.~F.~Welton and R.~Keller, Self-consistent, unbiased root-mean-square emittance analysis. \textit{Rev. Sci. Instrum.} \textbf{75} (2004) 1646.

\bibitem{Leitner2010} 
D.~Leitner, Ion beam properties and their diagnostics for ECR ion source injector systems, Proc. 14th Beam Instrumentation Workshop, Santa Fe, NM, USA, May 2010.

\bibitem{Humphries1990} 
S.~Humphries, \textit{Charged Particle Beams} (John Wiley \& Sons, New York, 1990).

\bibitem{WARP} 
J.-L.~Vay, P.~Colella, J.~W.~Kwan, P.~McCorquodale, D.~B.~Serafini, A.~Friedman, D.~P.\ Grote, G.\ Westenskow, J.-C.~Adam, A.~H\'eron and I.~Haber,
Application of adaptive mesh refinement to particle-in-cell simulations of plasmas and beams. \textit{Phys. Plasmas} \textbf{11} (2004) 2928.

\bibitem{Chauvin2012} 
N.~Chauvin, Space-Charge Effects, in these proceedings.

\bibitem{Todd2008} 
D.~S.~Todd, D.~Leitner and M.~Strohmeier, Low energy beam diagnostics at the VENUS ECR ion source, Proc. 13th Beam Instrumentation Workshop, Tahoe City, CA, May 2008.

\bibitem{Child1911} 
C.~D.~Child, Discharge from hot CaO. \textit{Phys. Rev.} \textbf{32} (1911) 492.

\bibitem{Pierce1940} 
J.~R.~Pierce, Rectilinear electron flow in beams. \textit{J. Appl. Phys.} \textbf{11} (1940) 548.

\bibitem{Bohm1949} 
D.~Bohm, Minimum ionic kinetic theory for a stable sheath, in \textit{The Characteristics of Electrical Discharges in Magnetic Fields}, Eds. A.~Guthrie and R.~K.~Wakerling (McGraw-Hill, New York, 1947).

\bibitem{Chapman80} 
B.~N.~Chapman, \textit{Glow Discharge Processes} (John Wiley \& Sons, New York, 1980).

\bibitem{Becker04a} 
R.~Becker, NIGUN: a two-dimensional simulation program for the extraction of H$^-$ ions. \textit{Rev. Sci. Instrum.} \textbf{75} (2004) 1723.

\bibitem{Becker04b} 
R.~Becker, Mathematical formulation and numerical modelling of the extraction of H$^-$ ions, Proc. 10th Int. Symp. on the Production and Neutralization of Negative Ions and Beams. \textit{AIP Conf. Proc.} \textbf{763} (2005) 194.

\bibitem{Kalvas2011} 
T.~Kalvas, O.~Tarvainen, H.~Clark, J.~Brinkley and J.~\"Arje, Application of 3D code IBSimu for designing an H$^-$/D$^-$ extraction system for the Texas A\&M facility upgrade, Proc. 2nd Int. Symp. on Negative Ions, Beams and Sources, Takayama, Japan. \textit{AIP Conf. Proc.} \textbf{1390} (2011) 439.

\bibitem{Kuo1996} 
T.~Kuo, D.~Yuan, K.~Jayamanna, M.~McDonald, R.~Baartman, P.~Schmor and G.~Dutto, On the development of a 15{\u}mA direct current H$^-$ multicusp source. \textit{Rev. Sci. Instrum.} \textbf{67} (1996) 1316.

\bibitem{Keller2002} 
R.~Keller, D.~Cheng, R.~DiGennaro, R.~A.~Gough, J.~Greer, K.~N.~Leung, A.~Ratti, J.\ Reijonen, R.~W.\ Thomae, T.~Schenkel, J.~W.~Staples, R.~Yourd, A.~Aleksandrov,
M.~P.~Stockli and R.~W.~Welton, Ion-source and low-energy beam-transport issues with the front-end systems for the Spallation Neutron Source. \textit{Rev. Sci. Instrum.} \textbf{73} (2002) 914.

\bibitem{Midttun2012} 
\O.~Midttun, T.~Kalvas, M.~Kronberger, J.~Lettry, H.~Pereira, C.~Schmitzer and R.~Scrivens, A new extraction system for the Linac4 H$^-$ ion source. \textit{Rev. Sci. Instrum.} \textbf{83} (2012) 02B710.

\bibitem{Kalvas2012} 
T.~Kalvas, R.~F.~Welton, O.~Tarvainen, B.~X.~Han and M.~P.~Stockli, Simulation of H$^-$ ion source extraction systems for the Spallation Neutron Source with IBSimu. \textit{Rev. Sci. Instrum.} \textbf{83} (2012) 02A705.

\bibitem{Transport} 
D.~C.~Carey, K.~L.~Brown and F.~Rothacker, Third-order TRANSPORT with MAD input -- a computer program for designing charged particle beam transport systems, FERMILAB-Pub-98/310, Fermi National Accelerator Laboratory, October 1998.

\bibitem{Cosy} 
K.~Makino and M.~Berz, COSY INFINITY Version 9. \textit{Nucl. Instrum Meth.} \textbf{A558} (2005) 346.

\bibitem{Dimad} 
R.~V.~Servranckx, Users' guide to the program DIMAD, TRI-DN-93-K233, Triumf Design Note, July 1993.

\bibitem{Gios} 
H.~Wollnik, B.~Hartmann and M.~Berz, Principles of GIOS and COSY. \textit{AIP Conf. Proc.} \textbf{177} (1988) 74.

\bibitem{Path} 
A.~Perrin, J.-F.~Amand, T.~M\"utze, J.-B.~Lallement and S.~Lanzone, Travel user manual, CERN, April 2007.

\bibitem{Trace} 
K.~R.~Crandall and D.~P.~Rusthoi, TRACE 3-D documentation, LA-UR-97-886, Los Alamos National Laboratory Report, May 1997.

\bibitem{Self1963} 
S.~A.~Self, Exact solution of the collisionless plasma-sheath equation. \textit{Phys. Fluids} \textbf{6} (1963) 1762.

\bibitem{Whealton77} 
J.~H.~Whealton, Optics of single-stage accelerated ion beams extracted from a plasma.
\textit{Rev. Sci. Instrum.} \textbf{48} (1977) 829.

\bibitem{Whealton78} 
J.~H.~Whealton, E.~F.~Jaeger and J.~C.~Whitson, Optics of ion beams of arbitrary perveance extracted from a plasma. \textit{J. Comput. Phys.} \textbf{27} (1978) 32.

\bibitem{PbGuns} 
J.~E.~Boers, PBGUNS: a digital computer program for the simulation of electron and ion beams on a PC, Proc. Int. Conf. on Plasma Sciences, Vancouver, BC, 7--9 June 1993.

\bibitem{Kobra} 
P.~Sp\"adtke, KOBRA3-INP user manual, 2000.

\bibitem{Kalvas2010} 
T.~Kalvas, O.~Tarvainen, T.~Ropponen, O.~Steczkiewicz, J.~\"Arje and H.~Clark, IBSimu: a three-dimensional simulation software for charged particle optics. \textit{Rev. Sci. Instrum.} \textbf{81} (2010) 02B703.

\bibitem{IBSimu_web} 
T.~Kalvas, Ion beam simulator, the distribution website of IBSimu code,
http://ibsimu.sourceforge.net.

\bibitem{Poisson_sf} 
K.~Halbach and R.~F.~Holsinger, SUPERFISH -- a computer program for evaluation of RF cavities with cylindrical symmetry. \textit{Part. Accel.} \textbf{7} (1976) 213.

\bibitem{FEMM} 
D.~C.~Meeker, Finite element method magnetics, Version 4.0.1 (3 December 2006 build), http://www.femm.info.

\bibitem{Radia} 
O.~Chubar, P.~Elleaume and J.~Chavanne, Radia3D -- a computer program for calculating static magnetic fields, http://www.esrf.eu/Accelerators/Groups/InsertionDevices/Software/Radia.

\bibitem{Cobham} 
Cobham Vector Fields, Opera, http://www.cobham.com.

\bibitem{Comsol} 
Comsol Multiphysics, http://www.comsol.com.

\bibitem{IES} 
Integrated Engineering Software, http://www.integratedsoft.com.

\end{thebibliography}
\end{document}